\documentclass[aip,amsmath,amssymb,reprint,floatfix]{revtex4-2}
\usepackage{graphicx}
\usepackage{dcolumn}
\usepackage{bm}
\usepackage[utf8]{inputenc}
\usepackage[T1]{fontenc}
\usepackage{mathptmx}
\usepackage{etoolbox}
\usepackage{placeins}
\usepackage{subfloat}
\usepackage{caption}
\usepackage{subcaption}
\usepackage{tabularx}
\usepackage{xcolor}
\makeatletter
\def\@email#1#2{%
 \endgroup
 \patchcmd{\titleblock@produce}
  {\frontmatter@RRAPformat}
  {\frontmatter@RRAPformat{\produce@RRAP{*#1\href{mailto:#2}{#2}}}\frontmatter@RRAPformat}
  {}{}
}%
\makeatother

\begin{document}

\title[]{Statistics of punctuation in experimental literature – the remarkable case of \textit{Finnegans Wake} by James Joyce}

\author{Tomasz Stanisz}
\affiliation{Complex Systems Theory Department, Institute of Nuclear Physics, Polish Academy of Sciences, ul. Radzikowskiego 152, 31-342 Kraków, Poland}
\author{Stanis{\l}aw Dro\.zd\.z}%
 \email{stanislaw.drozdz@ifj.edu.pl}
 \affiliation{Complex Systems Theory Department, Institute of Nuclear Physics, Polish Academy of Sciences, ul. Radzikowskiego 152, 31-342 Kraków, Poland}
 \affiliation{Faculty of Computer Science and Telecommunications, Cracow University of Technology, ul. Warszawska 24, 31-155 Kraków, Poland}
\author{Jaros{\l}aw Kwapie\'n}
\affiliation{Complex Systems Theory Department, Institute of Nuclear Physics, Polish Academy of Sciences, ul. Radzikowskiego 152, 31-342 Kraków, Poland}


\begin{abstract}
As the recent studies indicate, the structure imposed onto written texts by the presence of punctuation develops patterns which reveal certain characteristics of universality. In particular, based on a large collection of classic literary works, it has been evidenced that the distances between consecutive punctuation marks, measured in terms of the number of words, obey the discrete Weibull distribution -- a discrete variant of a distribution often used in survival analysis. The present work extends the analysis of punctuation usage patterns to more experimental pieces of world literature. It turns out that the compliance of the the distances between punctuation marks with the discrete Weibull distribution typically applies here as well. However, some of the works by James Joyce are distinct in this regard -- in the sense that the tails of the relevant distributions are significantly thicker and, consequently, the corresponding hazard functions are decreasing functions not observed in typical literary texts in prose. \textit{Finnegans Wake} -- the same one to which science owes the word \textit{quarks} for the most fundamental constituents of matter -- is particularly striking in this context. At the same time, in all the studied texts, the sentence lengths -- representing the distances between sentence-ending punctuation marks -- reveal more freedom and are not constrained by the discrete Weibull distribution. This freedom in some cases translates into long-range nonlinear correlations, which manifest themselves in multifractality. Again, a text particularly spectacular in terms of multifractality is \textit{Finnegans Wake}.
\end{abstract}

\maketitle

\begin{quotation}
Natural language has a number of traits that are often identified in complex systems -- like multilevel hierarchical organization, long-range correlations, and lack of characteristic scale, as evidenced by the presence of power laws along with fractal and multifractal structures. In fact, language is a system in which complexity is exhibited in an evident manner, as it combines relatively simple elements into structures capable of expressing an infinite range of concepts at arbitrary level of sophistication. Quantitative analyses of language -- including studies on statistical properties of texts -- are aimed at revealing statistical laws describing the measurable properties of language and the processes responsible for their occurrence. Understanding those processes might be helpful in linguistic research around questions that still remain unanswered, like the ones regarding the origins of language, its learning and representation in the human brain. It also has the potential to enhance the tools used in the field of natural language processing, nowadays strongly influenced by deep learning and, in particular, large language models (LLMs).
\end{quotation}

\section{Introduction}

Language maintains its structure by rules governing various aspects of its organization, like syntax, semantics, and phonology. However, a complete description of natural language in terms of a finite set of such rules and relationships is extremely difficult. On the one hand, linguistic rules are precise enough to make language an effective tool of communication between different individuals; on the other hand, they exhibit certain level of flexibility to allow for emergence of new forms. Also, they are often subject to exceptions.

Some insight into the structure of natural language can be gained by studying its written representation~\cite{Halliday1985} from a statistical perspective~\cite{Stanisz2024}. Such an approach allows one to identify certain quantitative relationships, pertaining to general, ``global'' properties of language. The observed relationships are often expressed in terms of linguistic laws. Famous examples of such laws include Zipf's law~\cite{Zipf1949,Piantadosi2014}, Heaps' law~\cite{Heaps1978,Egghe2007}, or Menzerath-Altmann law~\cite{Altmann1980,Milicka2014}. Various characteristics of language structure, usage, and evolution are investigated with the use of concepts and methods originating in information theory~\cite{Montemurro2011,Debowski2020,Debowski2020a}, time series analysis~\cite{AlvarezLacalle2006,Ausloos2012,Drozdz2016,Liu2023,Sanchez2023}, and network theory~\cite{Cancho2001,Gong2011,Cong2014,Kulig2017,Stanisz2019,Souza2023}. Some of these concepts are fundamentally important from the perspective of the field of natural language processing~\cite{Jurafsky2024}, which has recently made significant advances due to the development of computational systems like large language models (LLMs)~\cite{Shanahan2023}.

In written language, one of the mechanisms of keeping a specific type of organization is the usage of punctuation. Punctuation marks establish the division of a text into segments, with sentences being the most immediate units of partition. However, one can study the partition determined by all punctuation marks present in a text -- not only the ones that mark the end of a sentence. It can be argued that such a partition is also meaningful: punctuation marks serve as ``breaks'' in a text, facilitating comprehension and reading out loud, as well as removing ambiguity~\cite{ChafeW-1988a}. In fact, it has been shown recently on a set of literary texts in prose in seven European languages\cite{Stanisz2023} that, in terms of length, the sequences of words between consecutive punctuation marks can be considered to behave more regularly than sentences. More precisely, the lengths of such sequences measured by the number of words can be described by the so-called discrete Weibull distribution~\cite{WeibullW-1951a,Nakagawa1975}, while for the lengths of whole sentences, no specific form of distribution is observed. Those results encourage for further research aimed at determining to what degree the statistical regularities related to punctuation are influential and universal in written language. Of course, universality perhaps applies only to orthographically similar languages.

This work is focused on several literary texts in prose (novels), in which the usage of punctuation is, in a sense, unusual, or marked, as the linguistic literature refers to such cases.  Some of these texts owe that characteristic to the usage of the stream-of-consciousness narrative mode, which attempts to use written language to mimic the ``unstructured'' flow of thoughts through mind, and results in some punctuation marks missing and in the presence of long, often unfinished sentences. There are also a few novels in the studied set, in which the partition into sentences is disregarded completely; these are examples of an experimental narrative technique, in which a single sentence spans over the whole novel. Clearly, the punctuation usage patterns in texts written with the mentioned narrative methods are different -- at least in some aspects -- from the ones that could be considered ``usual'' or ``unmarked''. A question arises, how such intentionally original usage of punctuation marks perturbs the observed statistics of punctuation: whether it constitutes an exception from the relevant statistical laws or whether it remains within the regime determined by those laws. A related, more general question is how much variation is present in the quantitative characteristics of written language when it comes to different ways and styles of writing.

\section{Analyzed data}

Table \ref{tab:analyzed_books} contains the titles of the literary works analyzed in this work -- novels written with the use of certain narrative techniques that are considered experimental or in some way different form the standard characteristics of written language. However, they are still considered to be pieces of prose. The traits determining the originality of each novel are also listed in Table.

\begin{table*}[ht]
\caption{The set of novels analyzed in this study. Apart from the title, the author, and the original language, the narrative techniques that are related to the non-standard punctuation usage are given as well. Language abbreviations: EN -- English, PL -- Polish, DE -- German, ES -- Spanish, FR -- French.}
\begin{ruledtabular}
\begin{tabular}{ll}
\textbf{Title -- Author (Original language)} & \textbf{Features influencing punctuation}\\
\hline
\textit{As I Lay Dying} -- W. Faulkner (EN) & stream-of-consciousness \\
\textit{Bramy raju} -- J. Andrzejewski (PL) & one sentence covers almost entire book  \\
\textit{Der Auftrag} -- F. Dürrenmatt (DE) & each of 24 chapters is a single sentence \\
\textit{Finnegans Wake} -- J. Joyce (EN) & stream-of-consciousness \\
\textit{Pointed Roofs} -- D. Richardson (EN) & stream-of-consciousness \\
\textit{Rayuela} -- J. Cortázar (ES) & stream-of-consciousness \\
\textit{Solar Bones} -- M. McCormack (EN) & no partition into sentences \\
\textit{The Waves} -- V. Woolf (EN) & stream-of-consciousness \\
\textit{Ulysses} -- J. Joyce (EN) & stream-of-consciousness \\
\textit{Zone} -- M. Énard (FR) &  most of the chapters have no partition into sentences \\
\end{tabular}
\end{ruledtabular}
\label{tab:analyzed_books}
\end{table*}


Some of the books listed in Table \ref{tab:analyzed_books} (\textit{Pointed Roofs}, \textit{Rayuela}, and \textit{The Waves}) have already been analyzed before~\cite{Stanisz2023} despite the fact that they contain special narrative techniques, because they turned out to be similar to the novels associated with more regular styles of writing in terms of the distances between consecutive punctuation marks. Therefore, these novels can be used as examples of texts in which the ``usual'' distribution of distances between consecutive punctuation marks is preserved even though they are written in a style associated with bending certain rules of punctuation usage.

The analyzed texts were appropriately preprocessed prior to being subject to analysis. Preprocessing consisted of removing annotations, foreword, chapter list, and additional information from the publisher. Then the texts were transformed into time series; these series have been used to determine the studied characteristics of punctuation usage. In each such series, consecutive numbers represent word counts between consecutive ''breakpoints” in the text; a ''breakpoint” is defined as a place in the text marked by the presence of any of the following punctuation marks: full stop, question mark, exclamation mark, ellipsis, comma, dash, colon, semicolon, left and right bracket. While different approaches are possible, here full stops following abbreviations are omitted, as they can be considered inherent parts of the abbreviations they accompany, not indicative of a breakpoint in the text. A similar line of reasoning can be applied to punctuation marks residing inside words -- apostrophes or hyphens joining two or more words into one -- these punctuation marks are also not taken into account when identifying breakpoint. Sequences of punctuation marks occurring next to each other (like "?!" or "...!") are treated as a single punctuation mark -- each such sequence is assumed to introduce a single breakpoint (this effectively means that there are no zeroes in the resulting time series). As a technical note, it is worth mentioning that some punctuation marks exist in several variants, whose usage conventions might vary across texts (dash, having more than one possible length, is an example) -- this variety has been taken into account in the analysis (different variants of the same punctuation mark have been replaced with one, standardized variant). Punctuation marks not mentioned here have been omitted.

\section{Distances between punctuation marks}

For the purpose of investigating punctuation in texts from a quantitative perspective, one can assume that the distribution of punctuation marks in texts is the result of some random process; then studying punctuation properties comes down to studying the properties of that process. The following line of reasoning can be proposed. Let writing be a procedure in which the writer puts a punctuation mark after each word with probability $p$ and puts no mark with probability $1-p$. No distinction between different types of punctuation marks is made as they may all be treated as serving roughly the same purpose: to introduce ``breaks'' into the written text. Each choice can be considered as a Bernoulli trial with the probability of success $p$. If the choices are independent of each other, then the distances measured by the number of words between consecutive punctuation marks have geometric distribution with parameter $p$, which describes the number of trials $k$ until the first success in a sequence of independent Bernoulli trials. However, studying such distances in real-world texts or linguistic corpora, one can come to a conclusion that a more general distribution is needed in order to represent empirical data correctly.

The geometric distribution can be generalized by allowing for a relationship between the outcomes of consecutive trials. One of possible generalizations is the so-called discrete Weibull distribution -- a discrete variant of the Weibull distribution used in various fields, including survival analysis, weather forecasting, and study of textual data~\cite{Johnson1994,MillerJr.1998,Altmann2009}. The distribution has two parameters: $p \in (0,1)$ and $\beta>0$; its cumulative distribution function is given by\cite{Nakagawa1975}:
\begin{equation}
\mathcal{F}(k)=1 - \left( 1-p \right)^{k^\beta}.
\label{eq:Weibull_CDF}
\end{equation}
For $\beta = 1$, it becomes the geometric distribution with parameter $p$. The significance of the above generalization can be conveniently expressed in terms of the hazard function $h(k)$ expressing the conditional probability that the $k$th trial will result in a success provided that no success has occurred in the preceding $k-1$ trials:
\begin{equation}
h(k) = \frac{P(k)}{1-\mathcal{F}(k-1)},
\label{eq:hazard_function}
\end{equation}
where $P(k)$ denotes the probability mass function. In the case of the discrete Weibull distribution it becomes~\cite{PadgettWJ-1985a}:
\begin{equation}\label{eq:hazard_Weibull}
h(k) = \; 1-\left(1-p\right)^{k^{\beta} - (k-1)^{\beta}}.
\end{equation}
For $\beta > 1$, $h(k)$ is an increasing function: the probability of success increases with the number of preceding unsuccessful trials; for $\beta < 1$, it is the opposite. In the special case of $\beta = 1$, the hazard function is constant and the resulting geometric distribution is said to be \textit{memoryless}. Thus, when the discussed formalism is used with regard to punctuation, $\beta$ describes how the probability of putting a punctuation mark after a word depends on the number of words already written since the last punctuation mark. The other parameter, $p$, is the probability of putting a punctuation mark right after the first word following the last punctuation mark: $p = h(1)$.

\begin{figure}[h]
\subfloat[]{\includegraphics[width=0.495\linewidth]{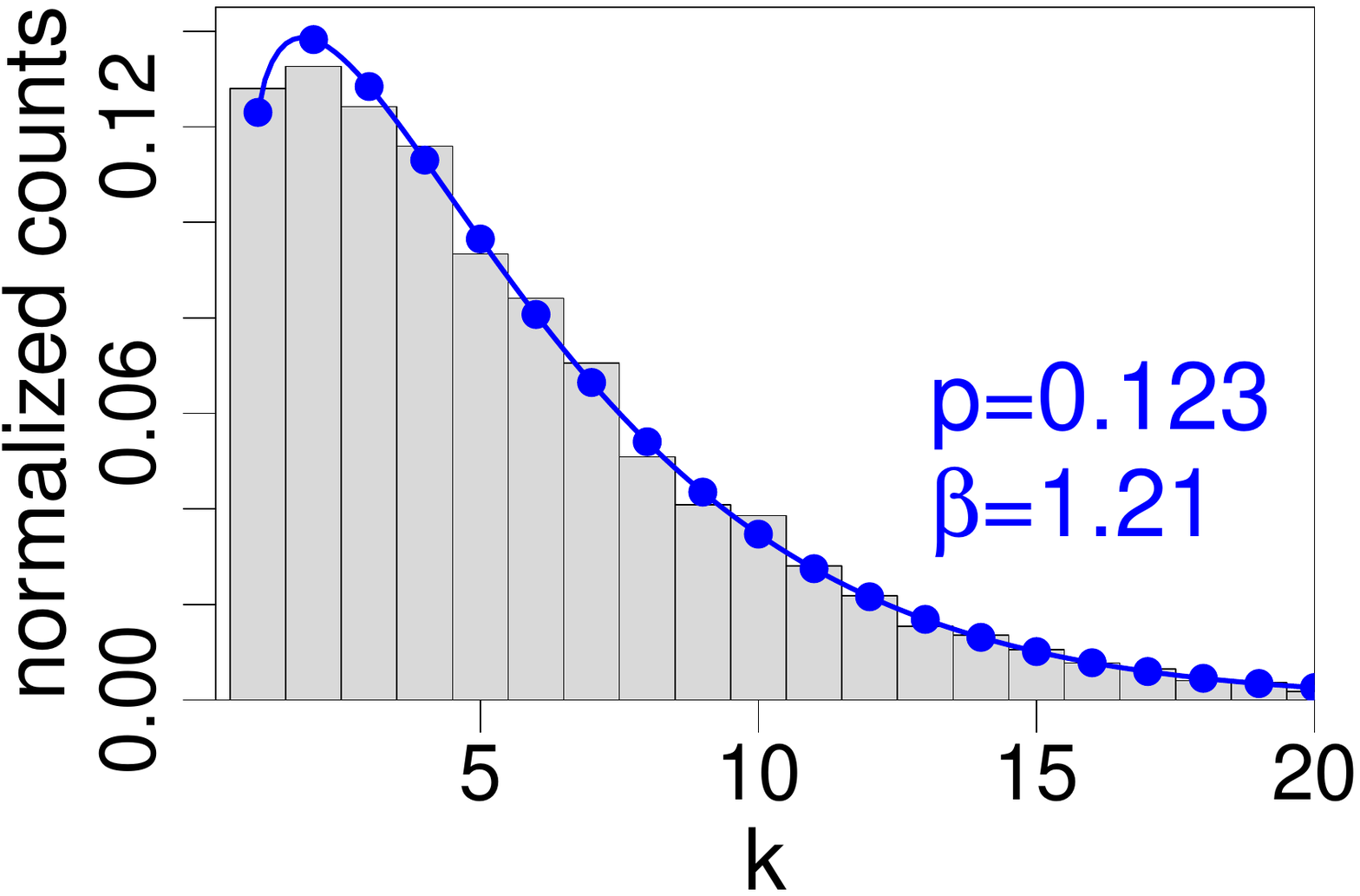}}
\subfloat[]{\includegraphics[width=0.495\linewidth]{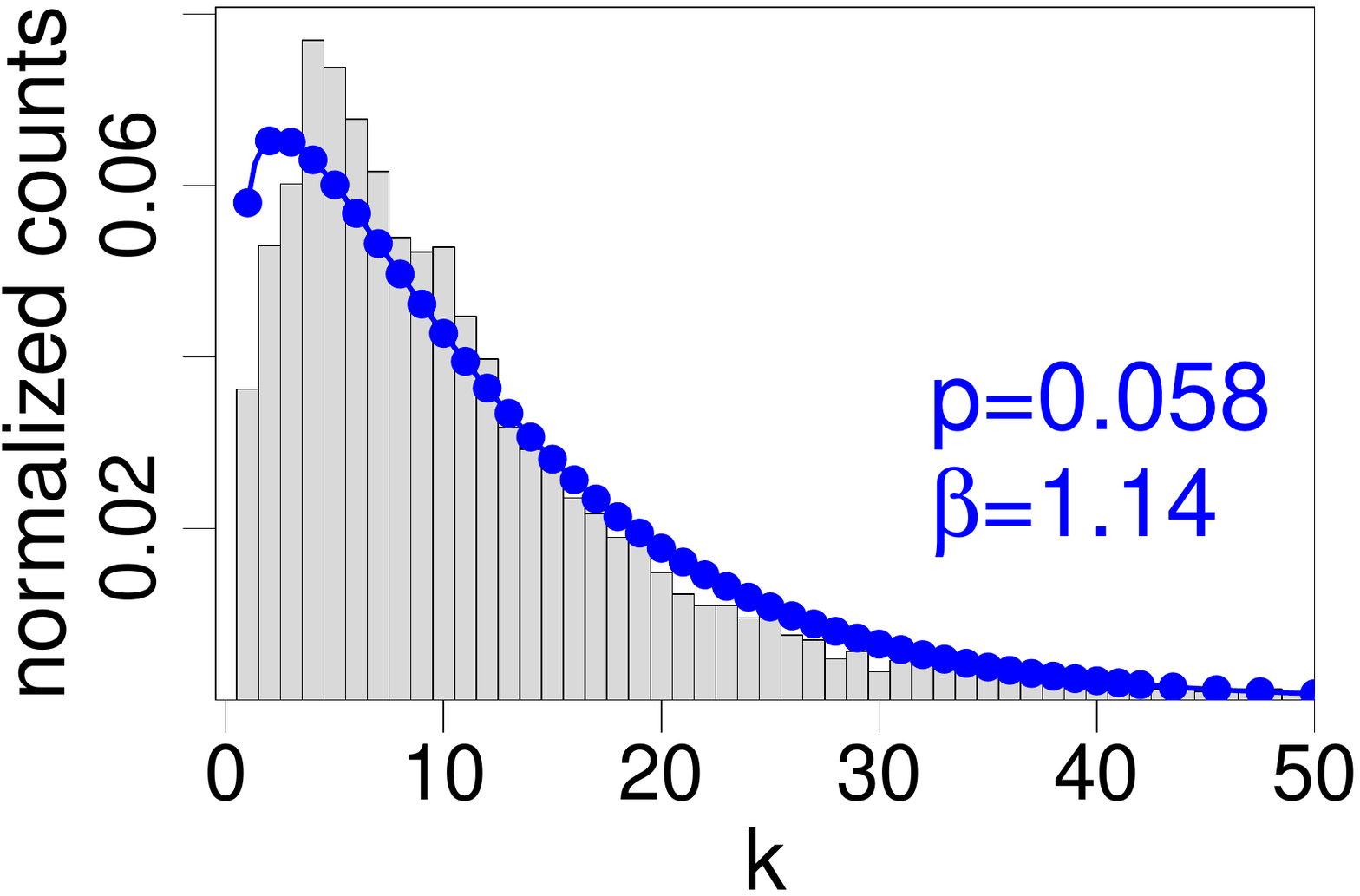}}

\vspace{0em}
\hfill
\subfloat[]{\includegraphics[width=0.495\linewidth]{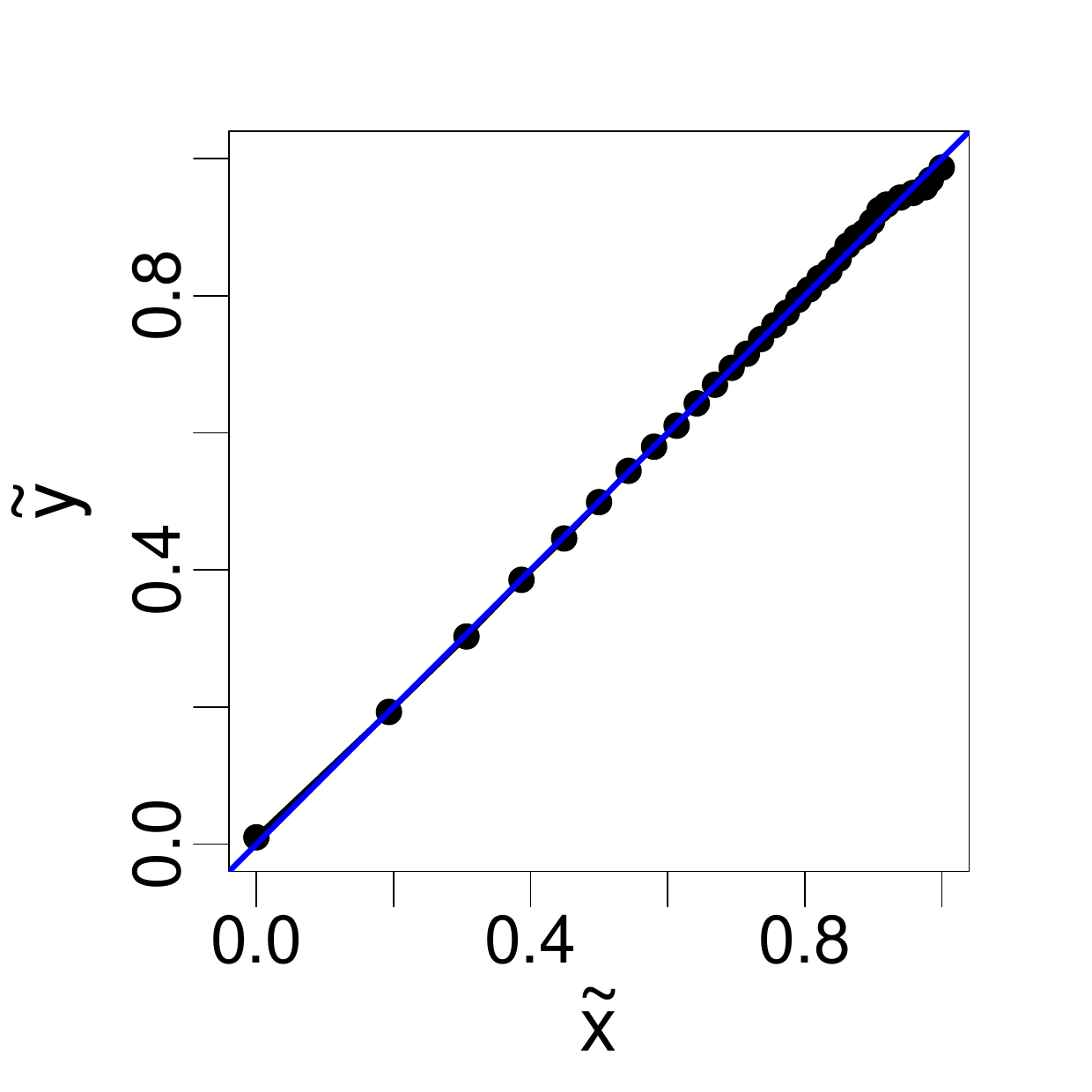}}
\subfloat[]{\includegraphics[width=0.495\linewidth]{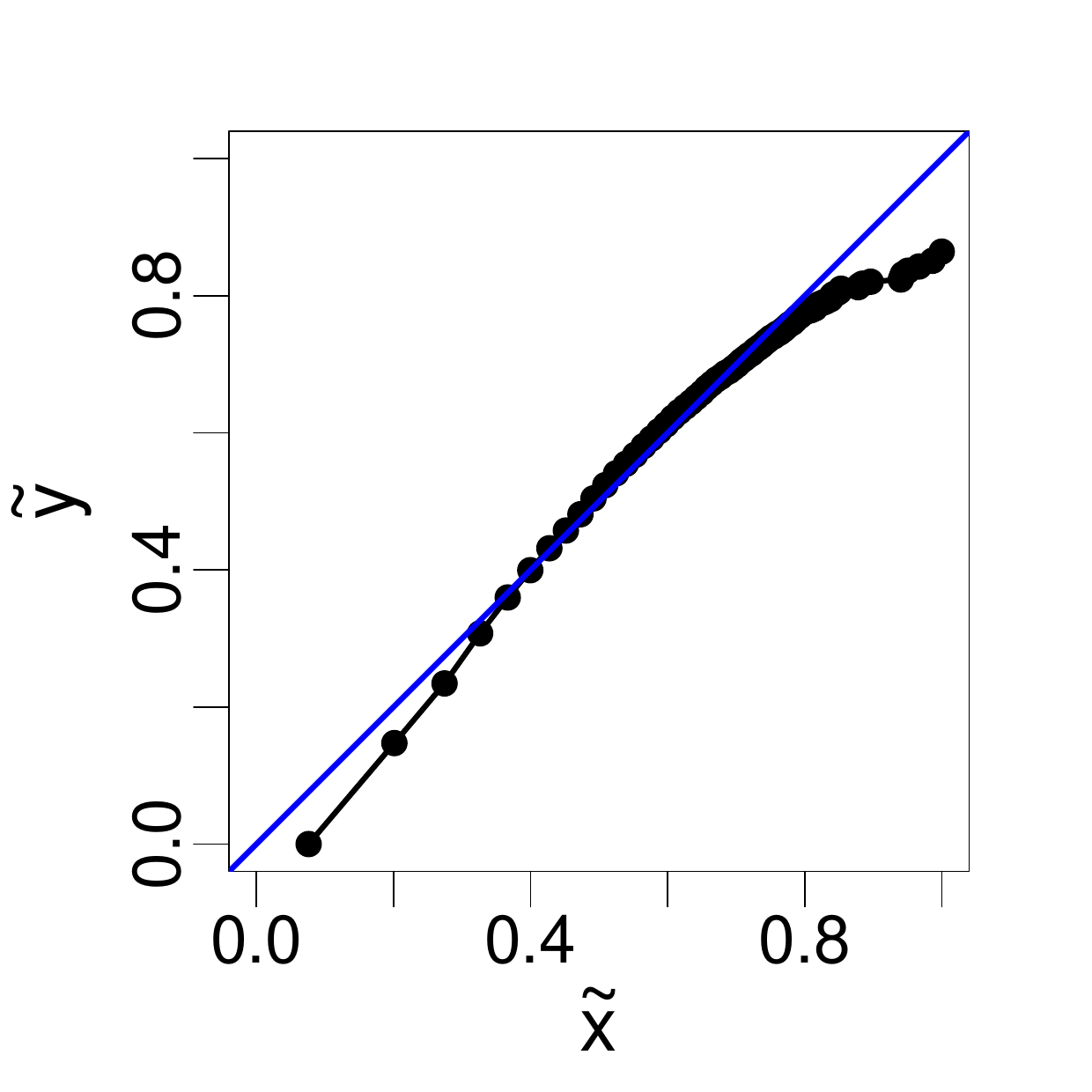}}

\vspace{0.75em}
\hfill
\subfloat[]{\includegraphics[width=0.495\linewidth]{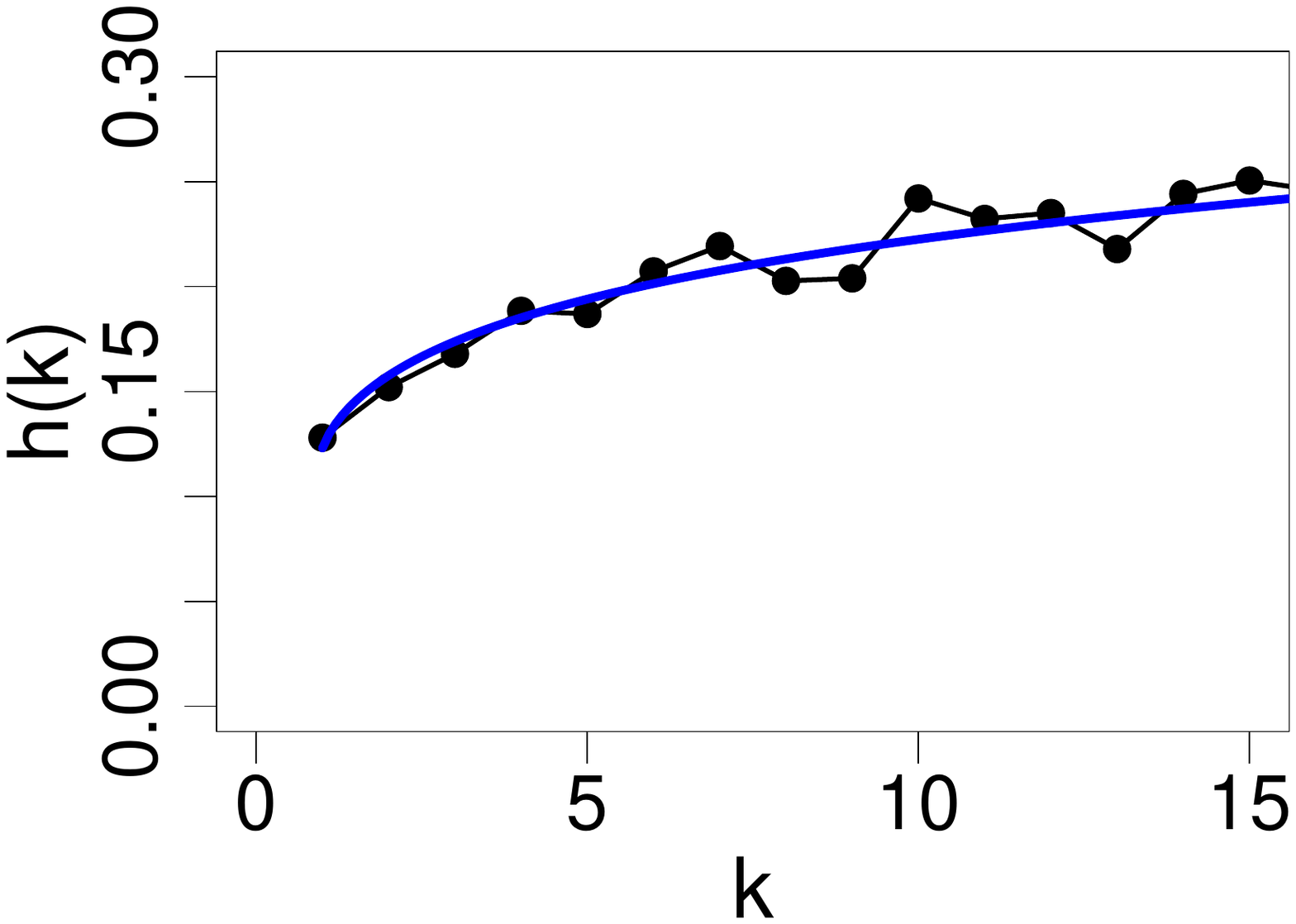}}
\subfloat[]{\includegraphics[width=0.495\linewidth]{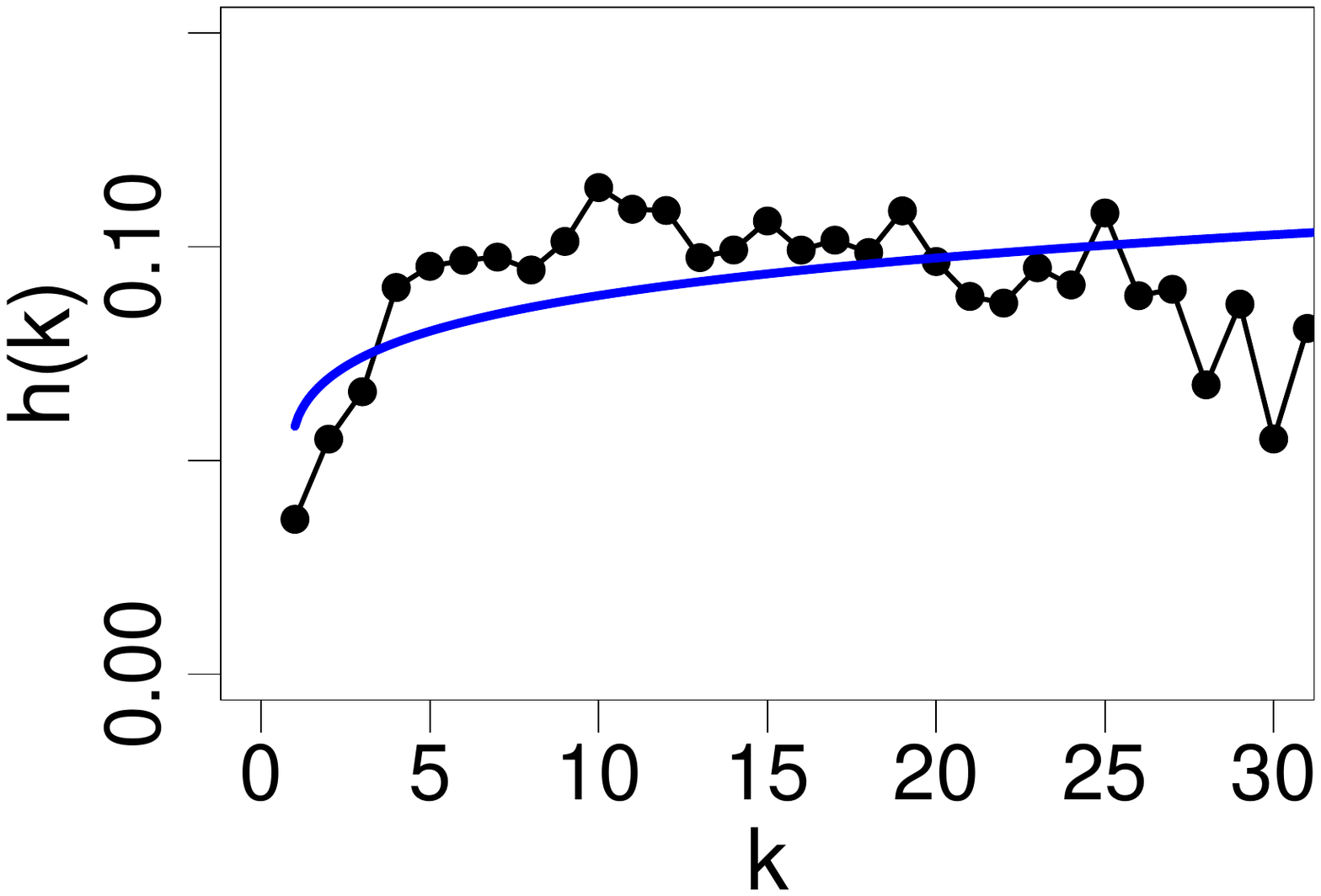}}
\vspace{1em}
\caption{Modeling the distribution of the distances between consecutive punctuation marks in \textit{Brave New World} by Aldous Huxley with the discrete Weibull distribution: all punctuation marks (left column) and sentence-ending marks only (right column). The latter plots are equivalent to the distribution of sentence lengths. The rows show: (top) the empirical distributions shown as gray histograms together with the fitted discrete Weibull distributions denoted by blue symbols, (middle) the rescaled Weibull plots with blue lines corresponding to the fitted distributions, and (bottom) the hazard functions for the empirical data -- marked in black -- and for the fitted discrete Weibull distributions -- marked in blue.}
\label{fig:Brave_new_world_analysis}
\end{figure}

A practical way of assessing and visualizing how well a given data set fits the Weibull distribution is to construct a so-called Weibull plot. For this purpose, Eq.~(\ref{eq:Weibull_CDF}) is rewritten in the form:
\begin{equation}
\log \left( -\log \left( 1-\mathcal{F}(k) \right) \right) = \beta \log k + \log \left( -\log \left( 1-p \right) \right).
\end{equation}
Then, for data originating from the discrete Weibull distribution with parameters $(p,\beta)$, a plot of the empirical cumulative distribution function $\mathcal{F}_{\rm emp}(k)$ in coordinates $(x,y)$ such that
\begin{equation}
\begin{aligned} 
&x=\log k \\
&y=\log\left( -\log \left( 1-\mathcal{F}_{\rm emp}(k) \right) \right),
\end{aligned}
\label{eq:Weibull_plot}
\end{equation}
results in a straight line with slope $\beta$ and intercept $\log\left( -\log \left( 1-p \right) \right)$. To make a comparison between the fits to different Weibull distributions easier, one can use the \textit{rescaled Weibull plot} with the coordinates rescaled linearly to $(\widetilde{x}, \widetilde{y})$, in which the plot fits the square $[0,1]\!\times\![0,1]$ and the reference line has slope 1 and intercept 0 (see Appendix~\ref{appendix_Weibull_plot} for the exact formulas describing the coordinate transformation). The advantage of using the rescaled Weibull plot is that a deviation of the data from some Weibull distribution corresponds to a deviation from the line $\widetilde{y} = \widetilde{x}$. An example of how the discrete Weibull distribution and the Weibull plots are applied to describe the distribution of punctuation in sample empirical data (the novel \textit{Brave New World} by Aldous Huxley) is presented in Fig.~\ref{fig:Brave_new_world_analysis}.

Fig.~\ref{fig:histograms_and_h} shows the distributions of the distances between consecutive punctuation marks (of any type) and the corresponding hazard functions for the novels listed in Tab.~\ref{tab:analyzed_books}. As mentioned before, these novels are regarded here as examples of the texts with some form of experimental literary style. By looking at the rescaled Weibull plots shown as the insets in the left column of Fig.~\ref{fig:histograms_and_h}, one can conclude that the level of agreement of the empirical distributions with the discrete Weibull distribution varies among the texts. While there are texts which keep their inter-mark distances within the regime determined by the discrete Weibull distribution despite a non-standard usage of punctuation (by disregarding a partition into sentences, for example), there are several texts, in which the form of the discussed distribution is different. However, the departure from the model distribution in some texts is caused by a few outlying observations only. This is especially evident for \textit{As I Lay Dying} and \textit{Ulysses} as both novels contain two disproportionately long sequences of words with no punctuation marks in between. By removing these sequences from the analyzed data, the agreement with the discrete Weibull distribution improves, although some discrepancy is still retained.

\begin{figure*}[!p]
\centering
\includegraphics[width=0.35\linewidth]{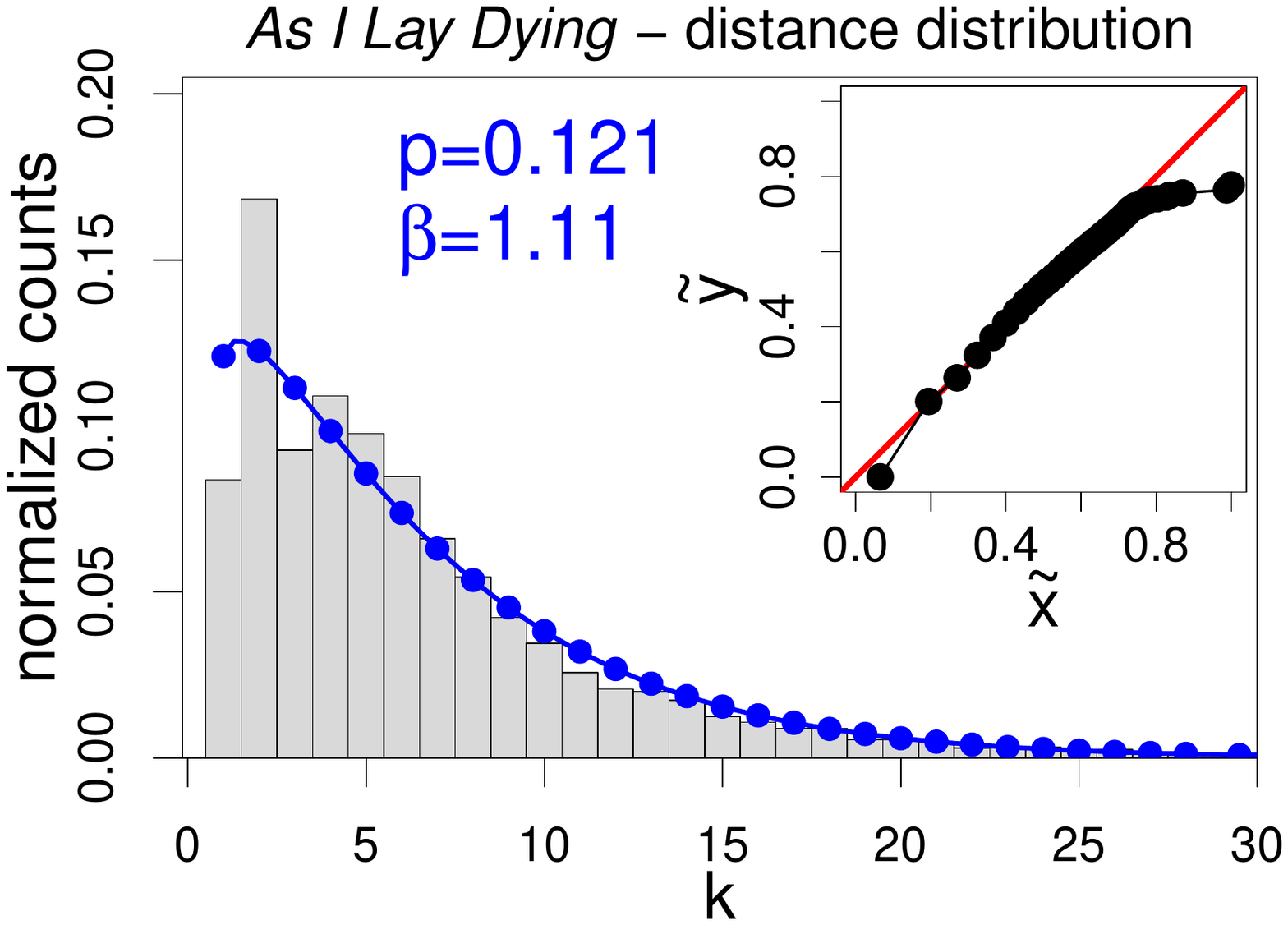}
\qquad
\includegraphics[width=0.35\linewidth]{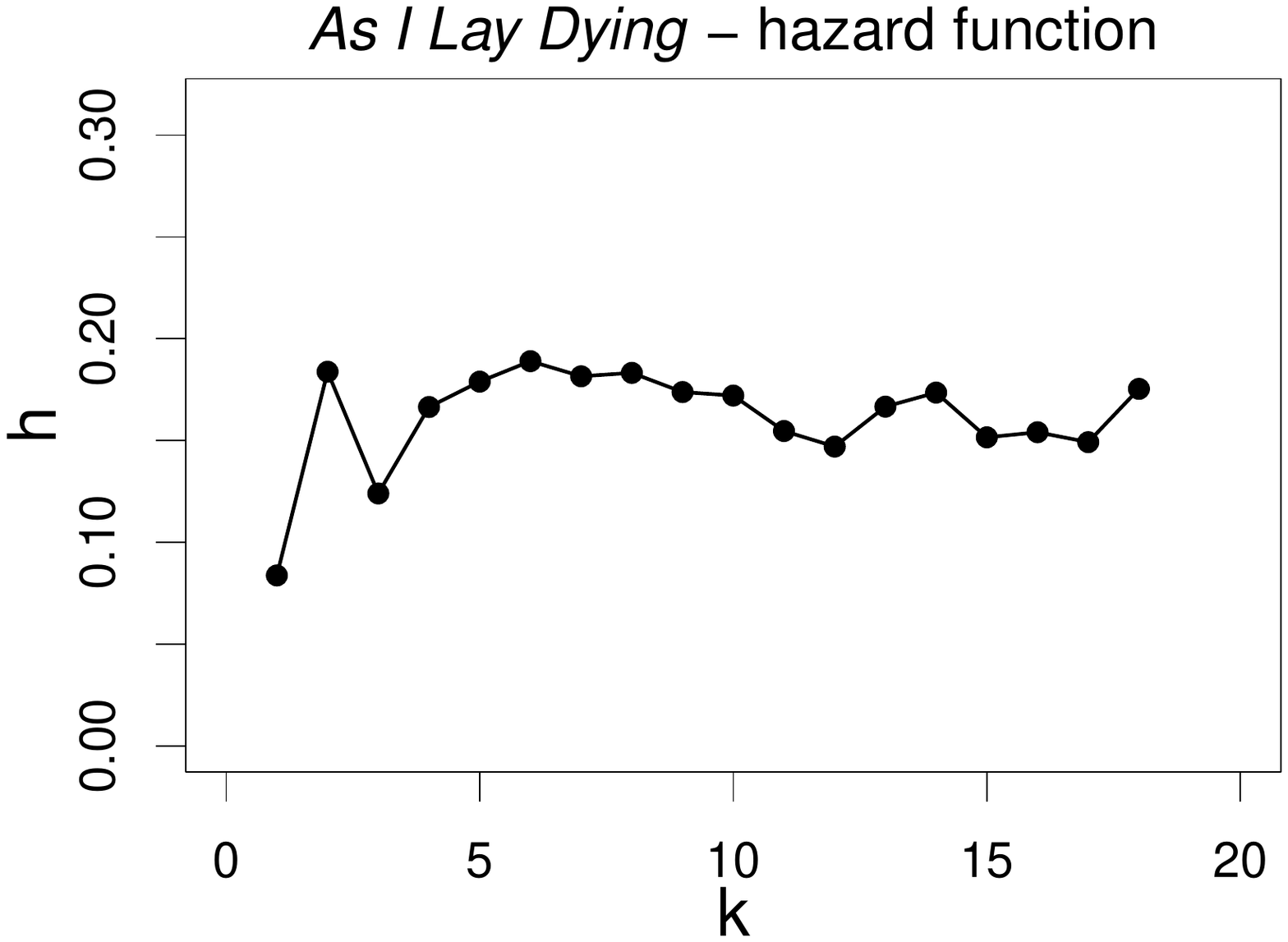}

\includegraphics[width=0.35\linewidth]{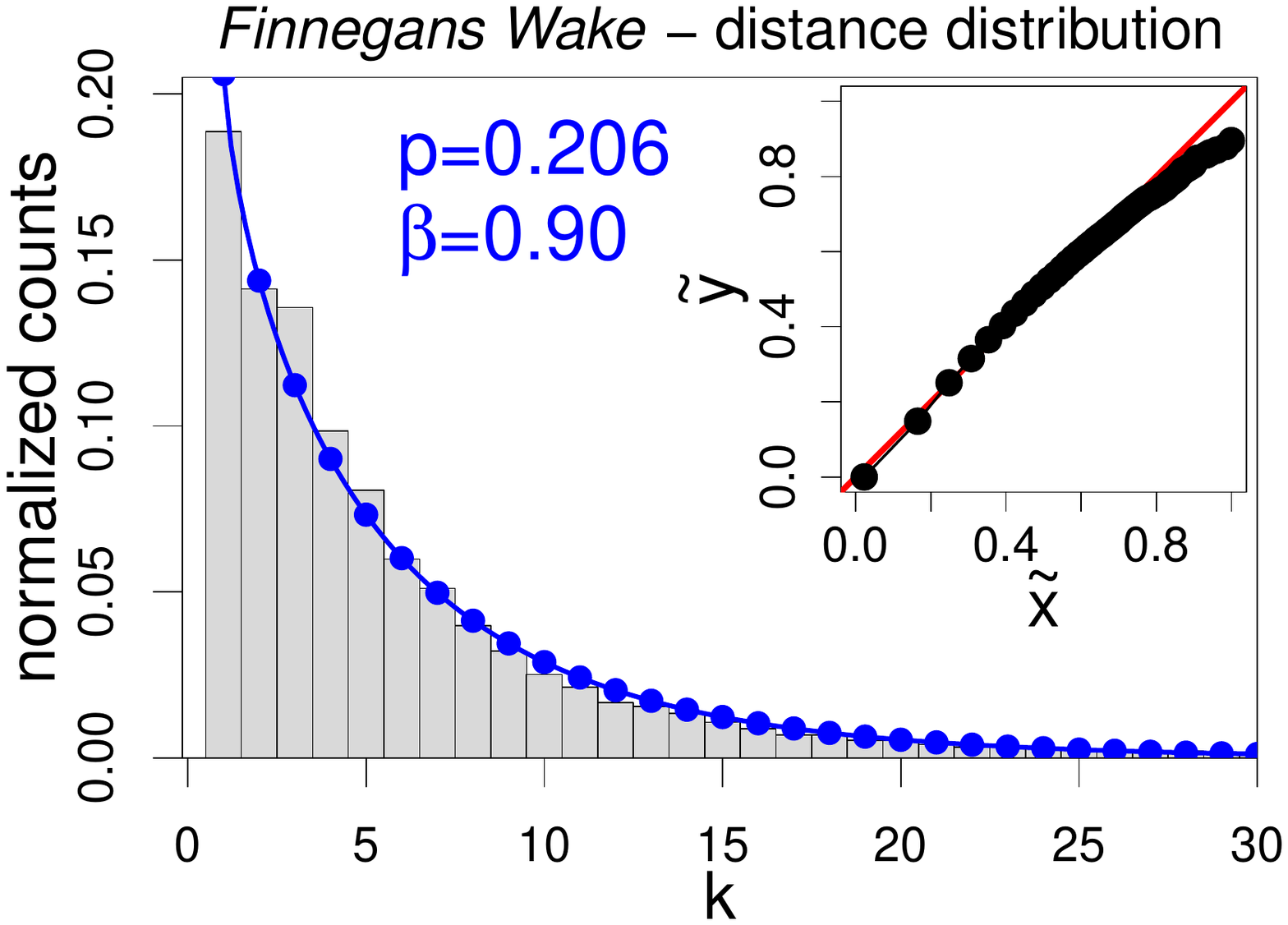}
\qquad
\includegraphics[width=0.35\linewidth]{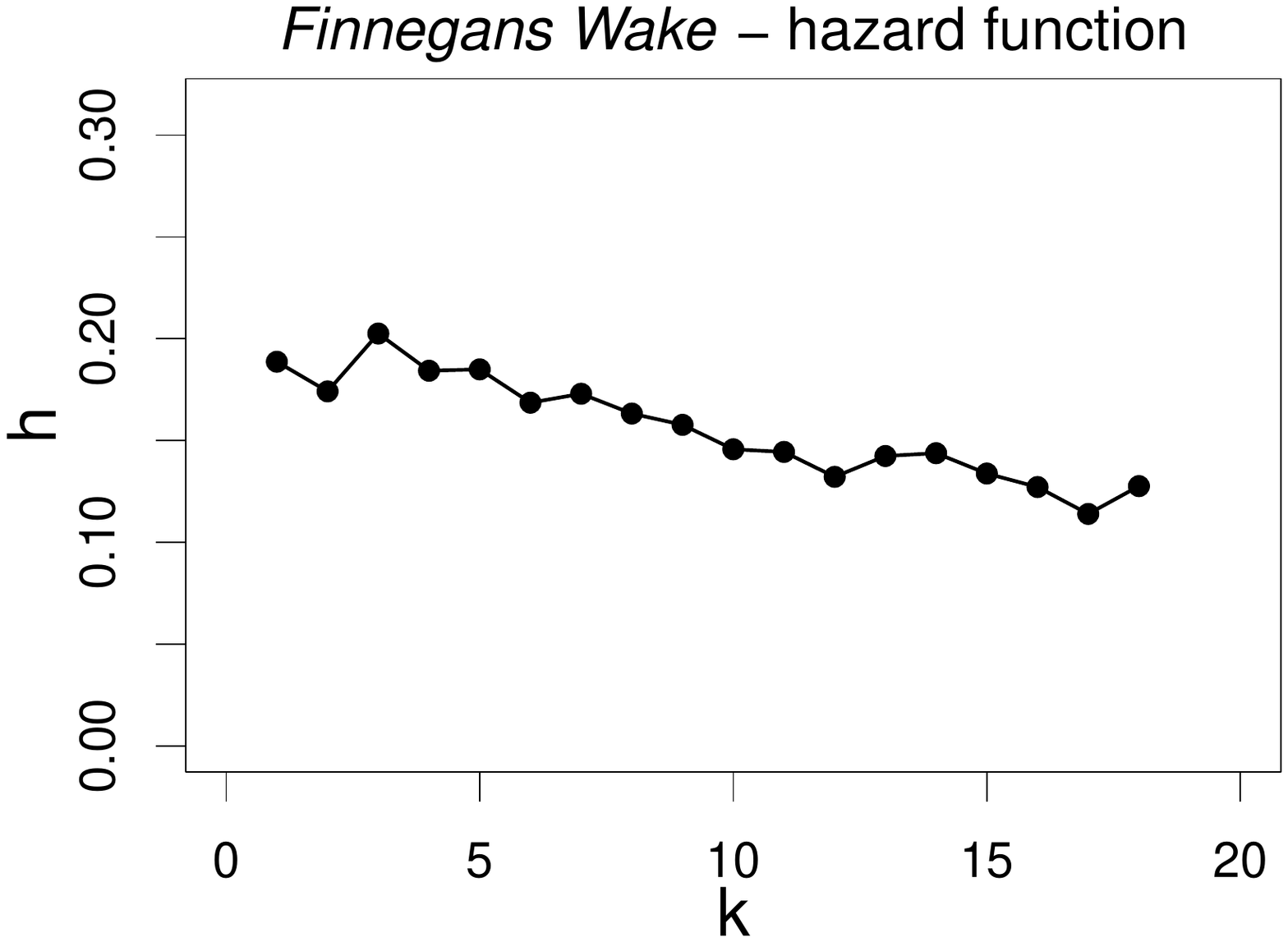}

\includegraphics[width=0.35\linewidth]{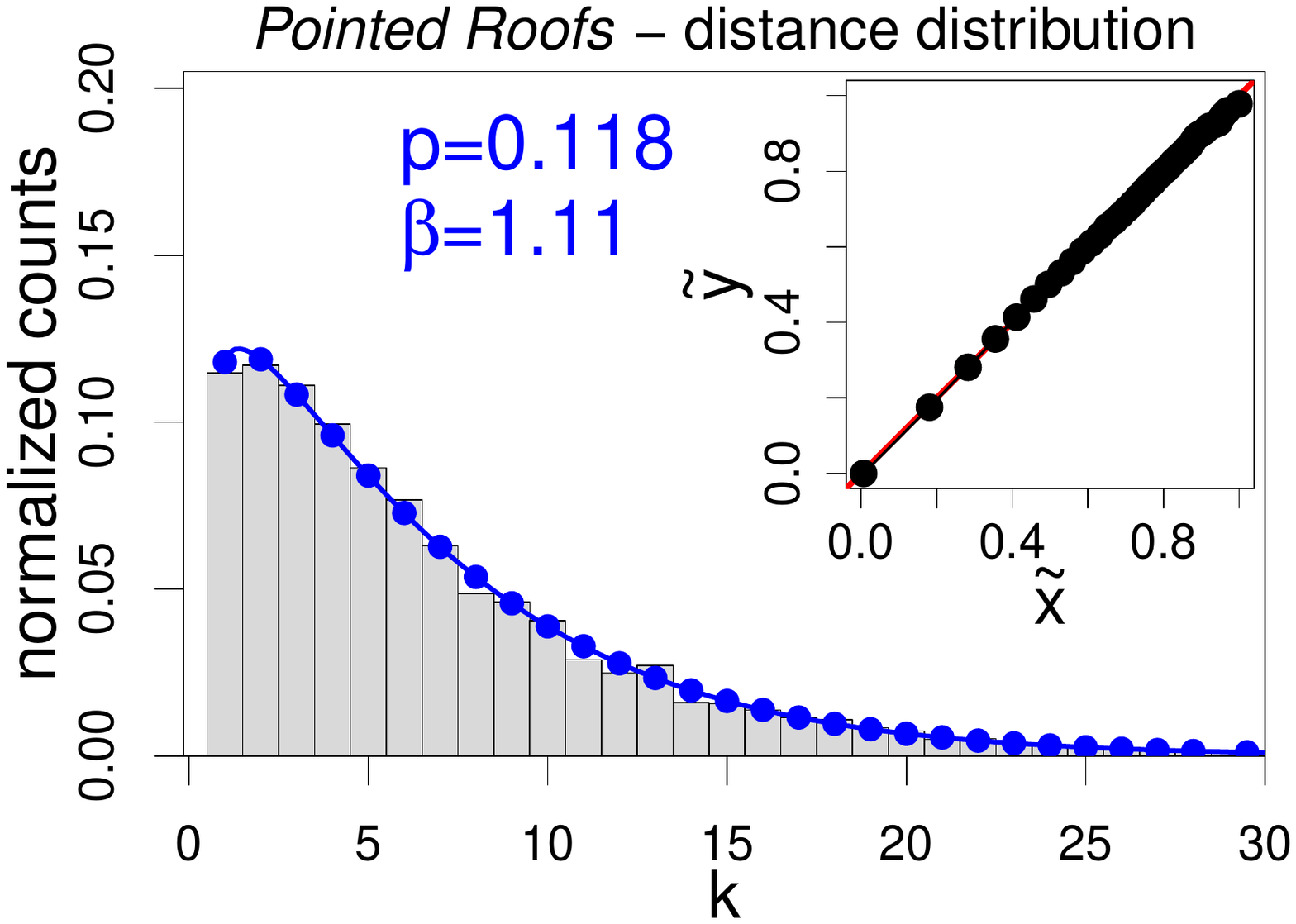}
\qquad
\includegraphics[width=0.35\linewidth]{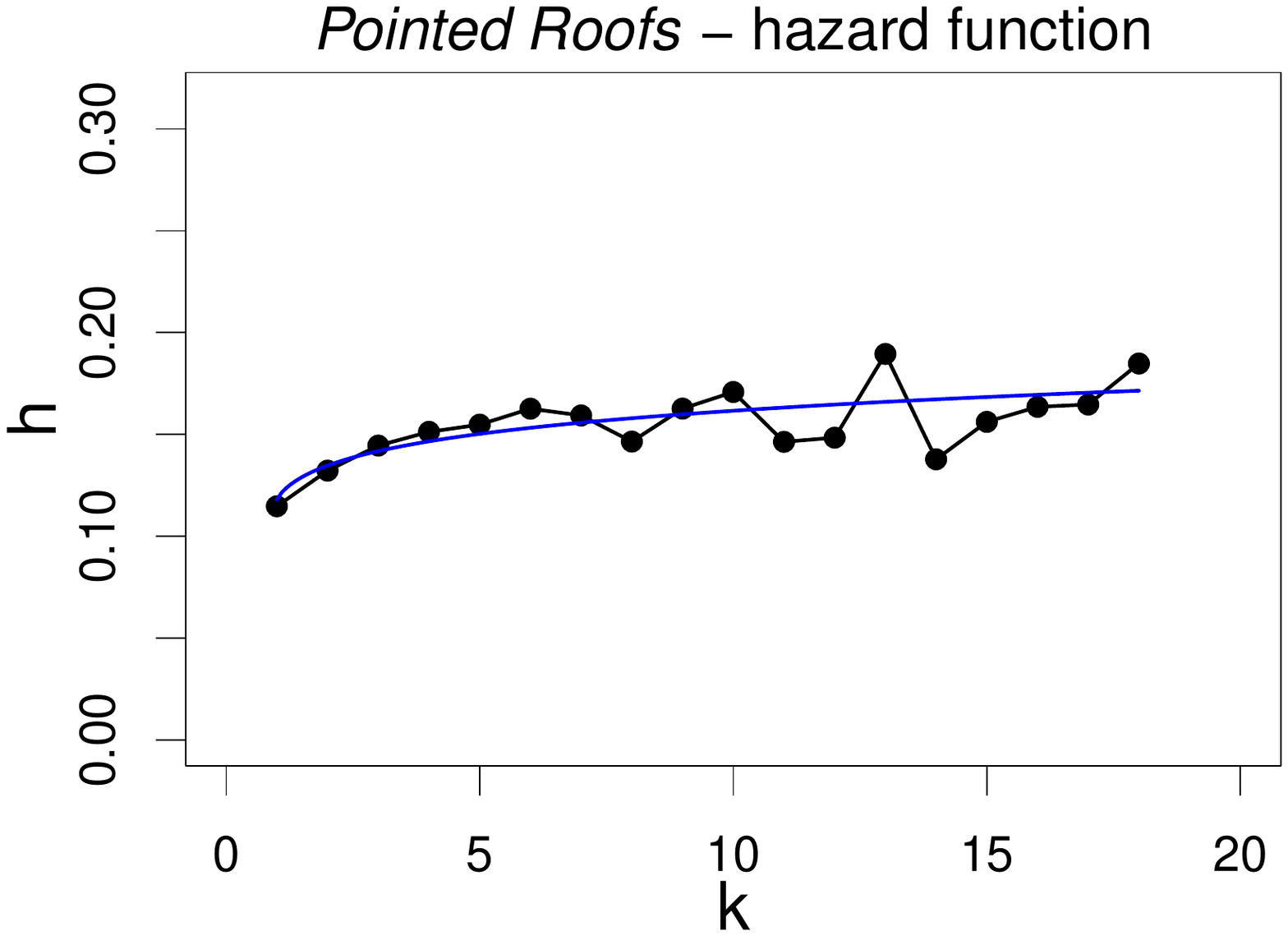}

\includegraphics[width=0.35\linewidth]{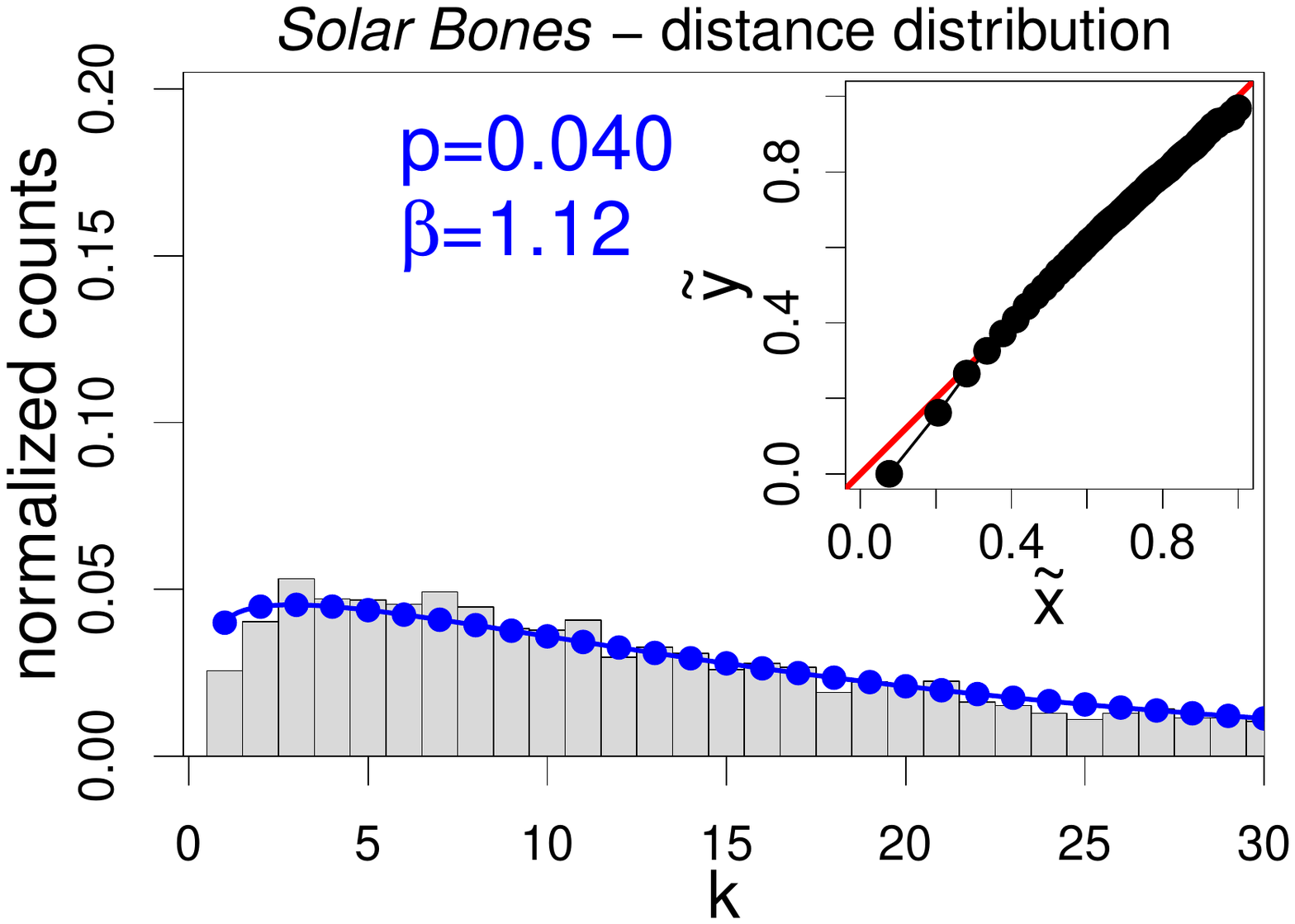}
\qquad
\includegraphics[width=0.35\linewidth]{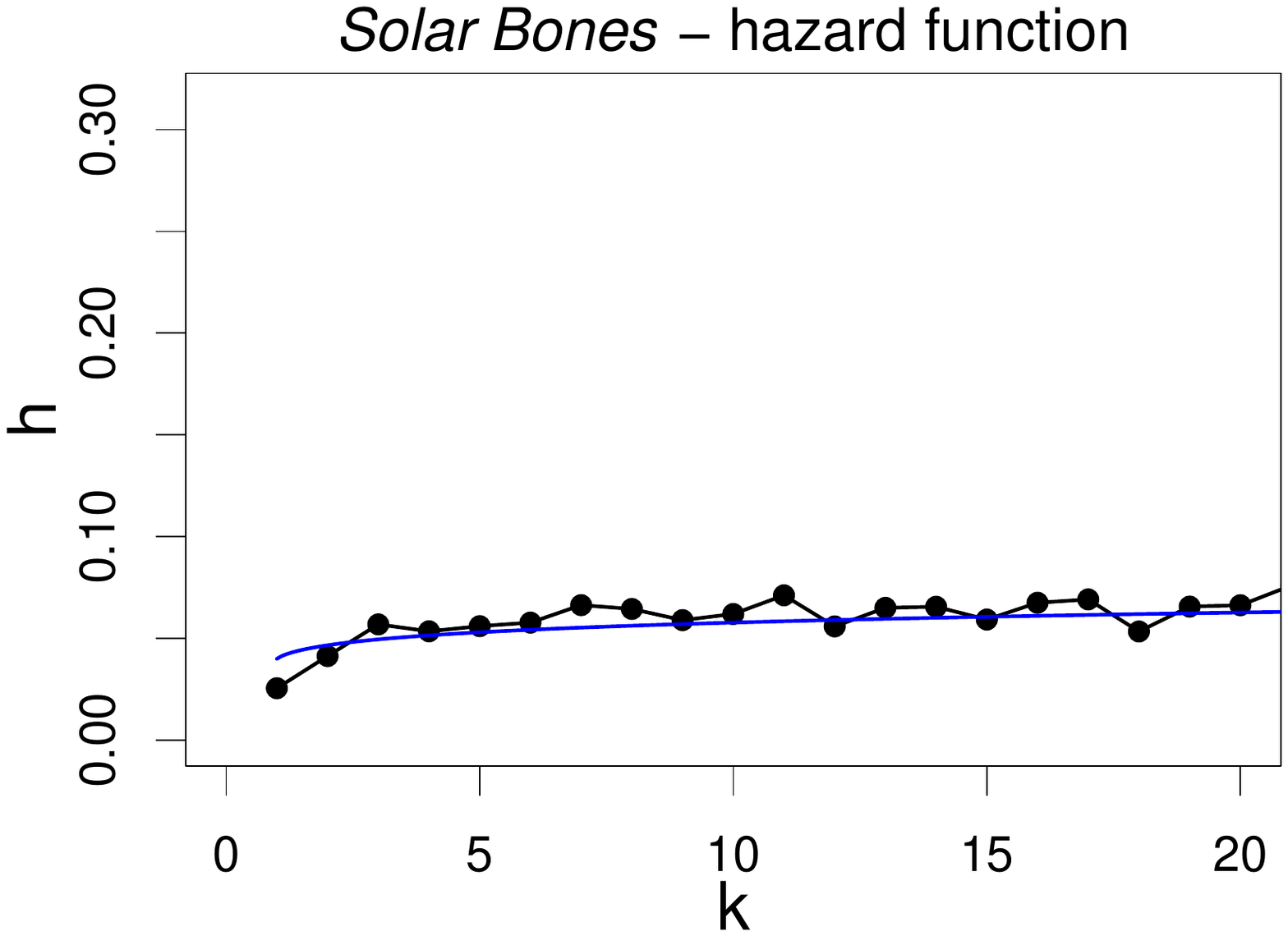}

\includegraphics[width=0.35\linewidth]{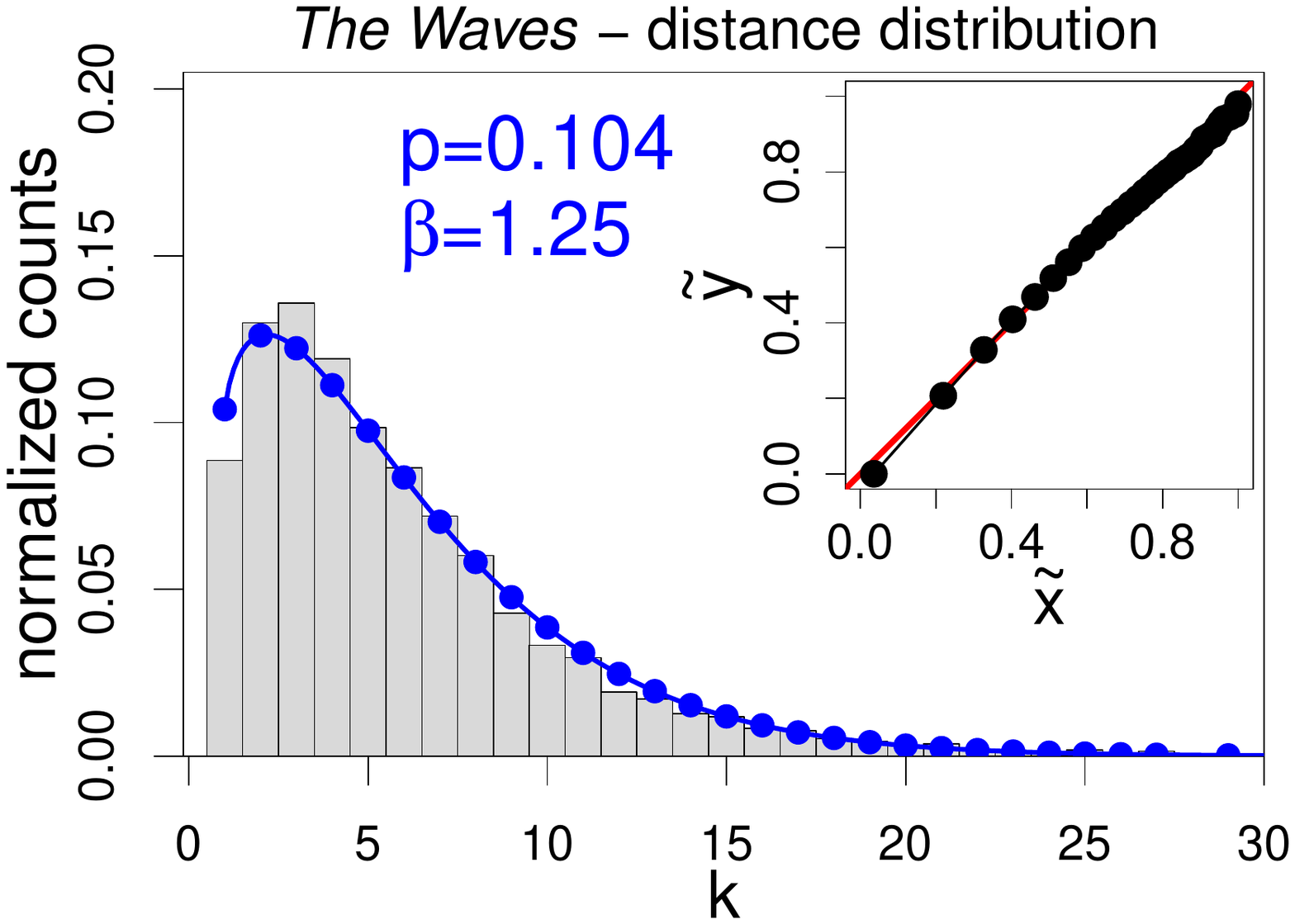}
\qquad
\includegraphics[width=0.35\linewidth]{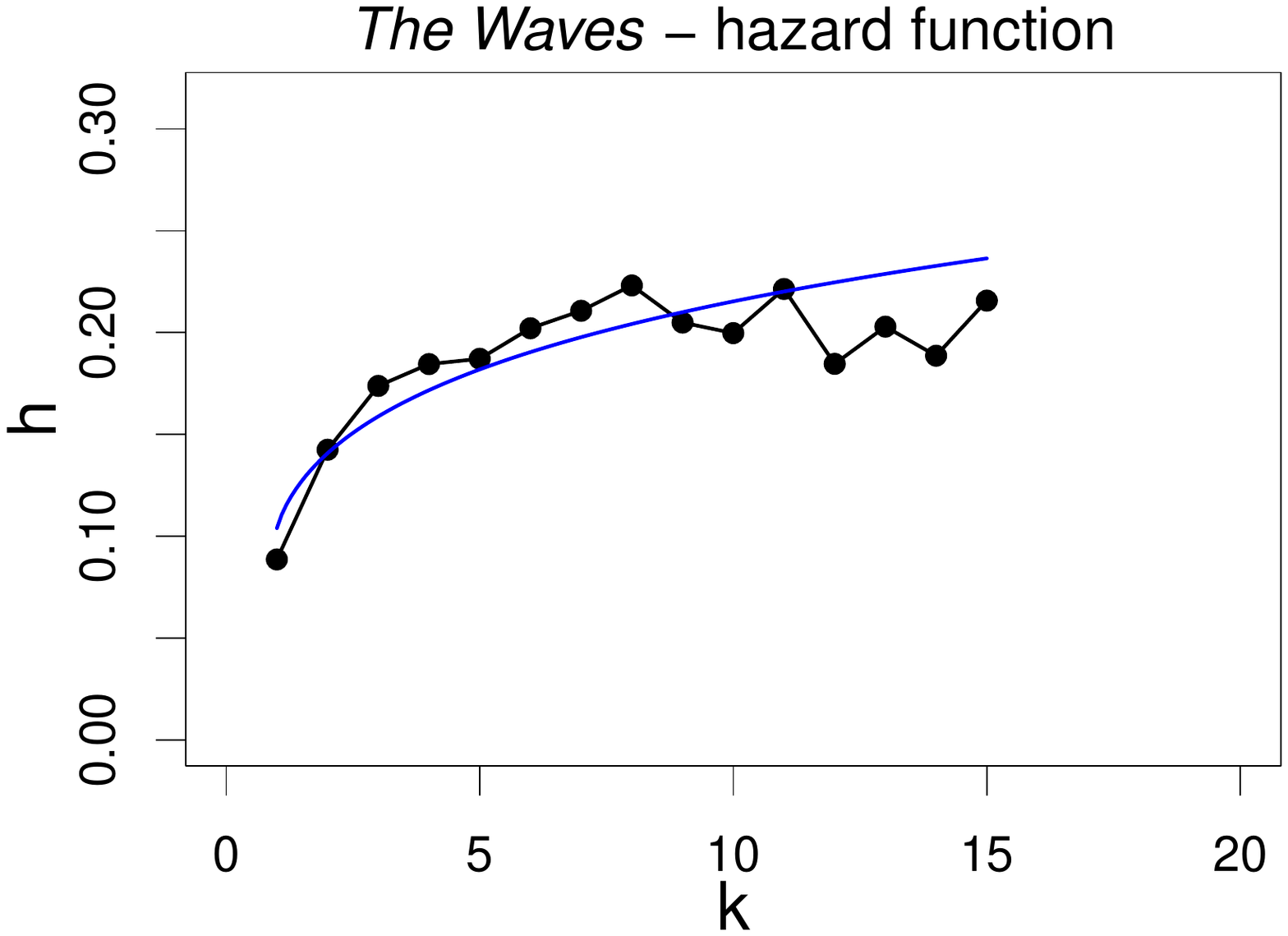}
\caption{Left column: (main) histograms of empirical inter-punctuation-mark distance distributions, along with the fitted discrete Weibull distributions, marked with blue symbols; (insets) the corresponding rescaled Weibull plots. Right column: the empirical hazard functions (black dots) and the hazard functions of the fitted discrete Weibull distributions if such fits are possible (blue curves). Each row corresponds to a particular novel.}
\label{fig:histograms_and_h}
\end{figure*}

\begin{figure*}[!p]
\ContinuedFloat
\includegraphics[width=0.35\linewidth]{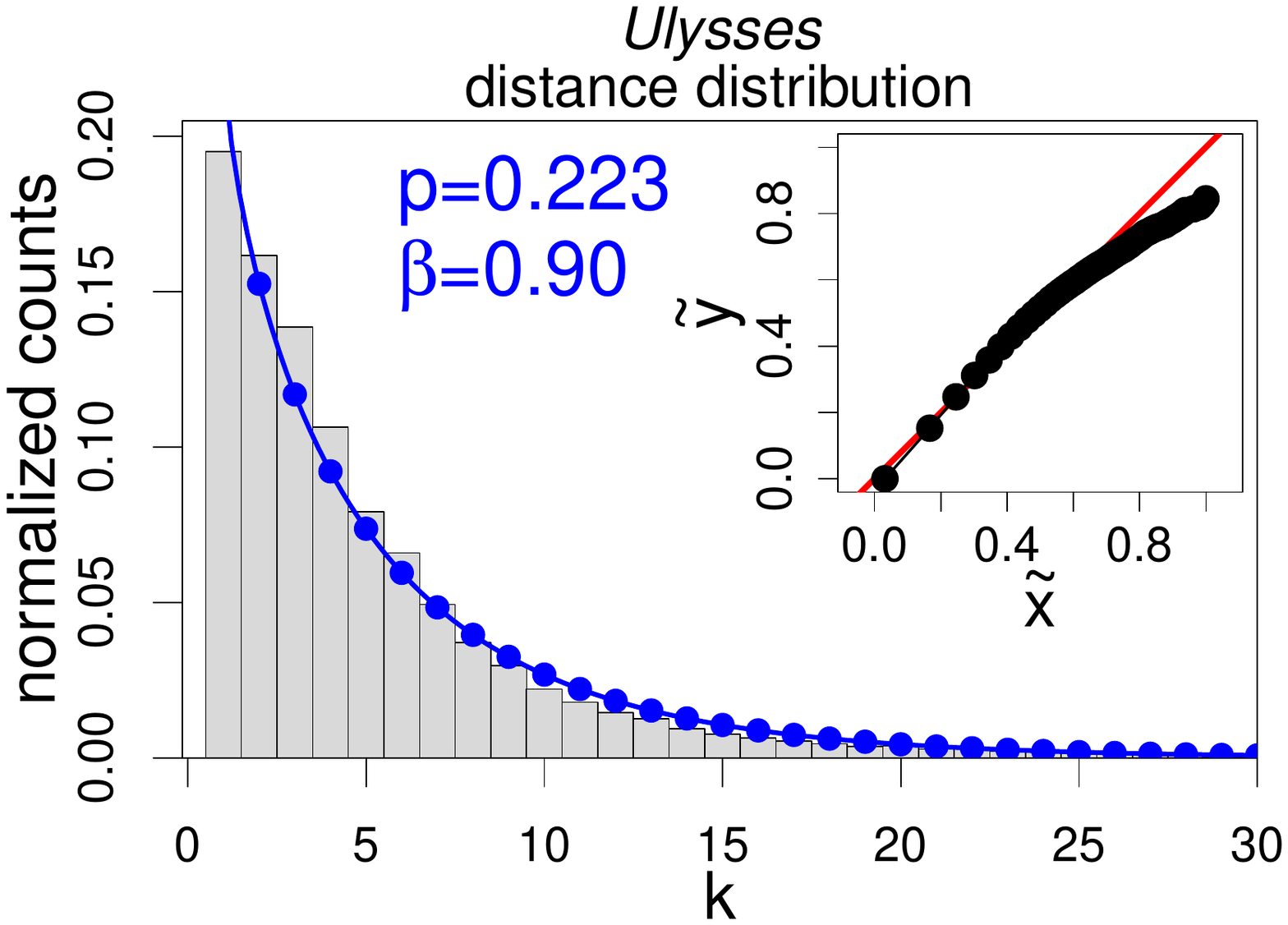}
\qquad
\includegraphics[width=0.35\linewidth]{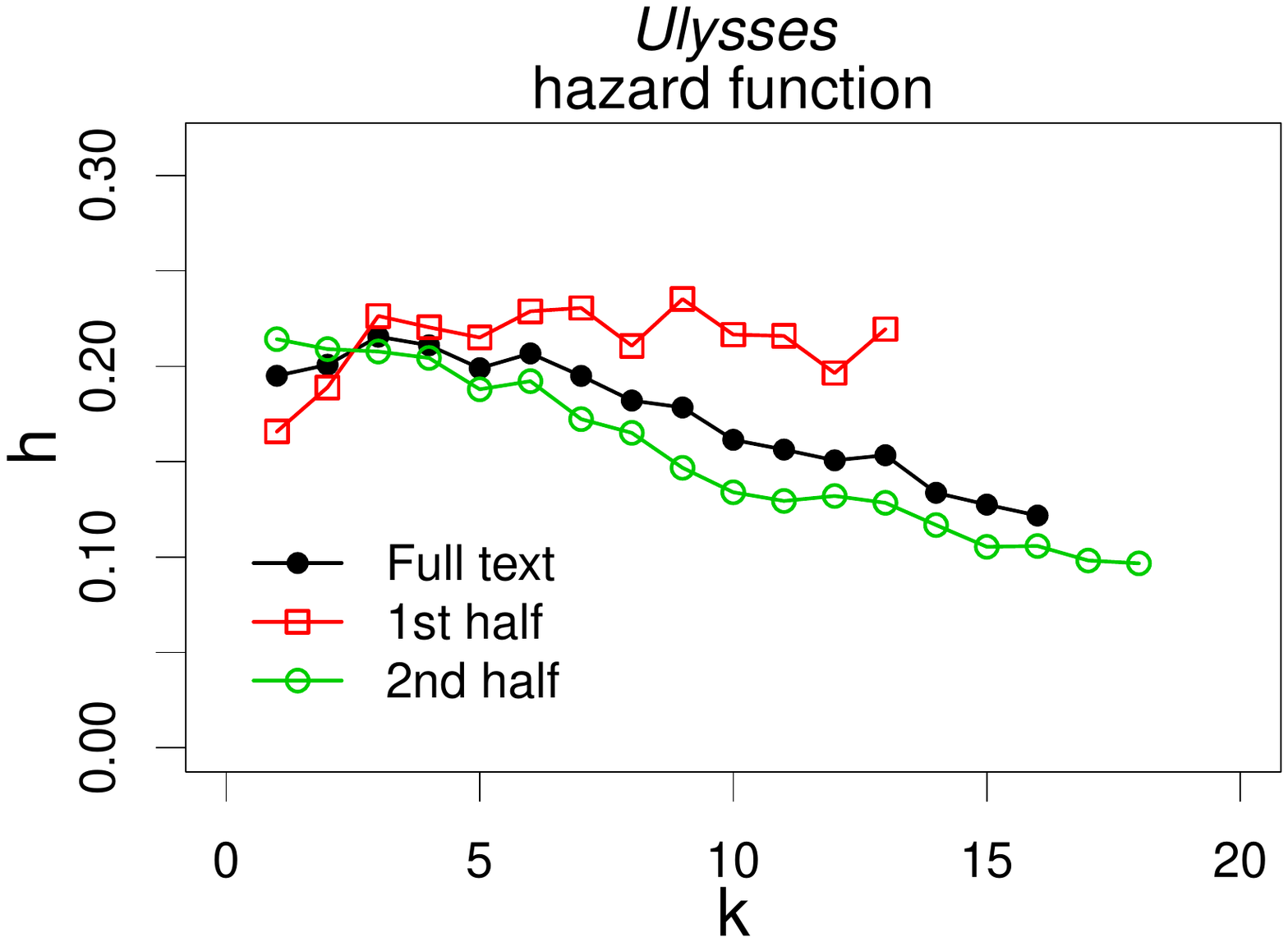}

\includegraphics[width=0.35\linewidth]{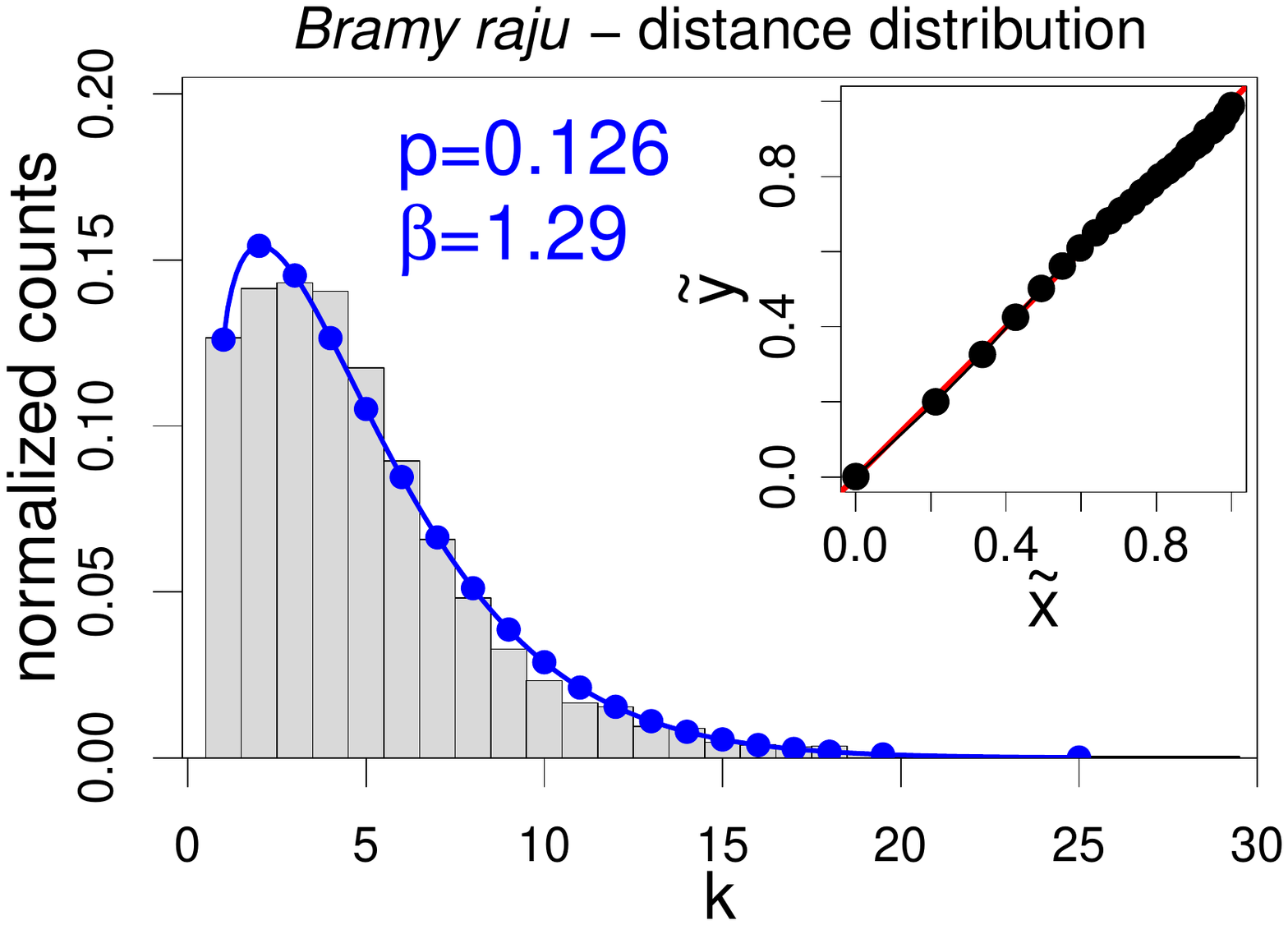}
\qquad
\includegraphics[width=0.35\linewidth]{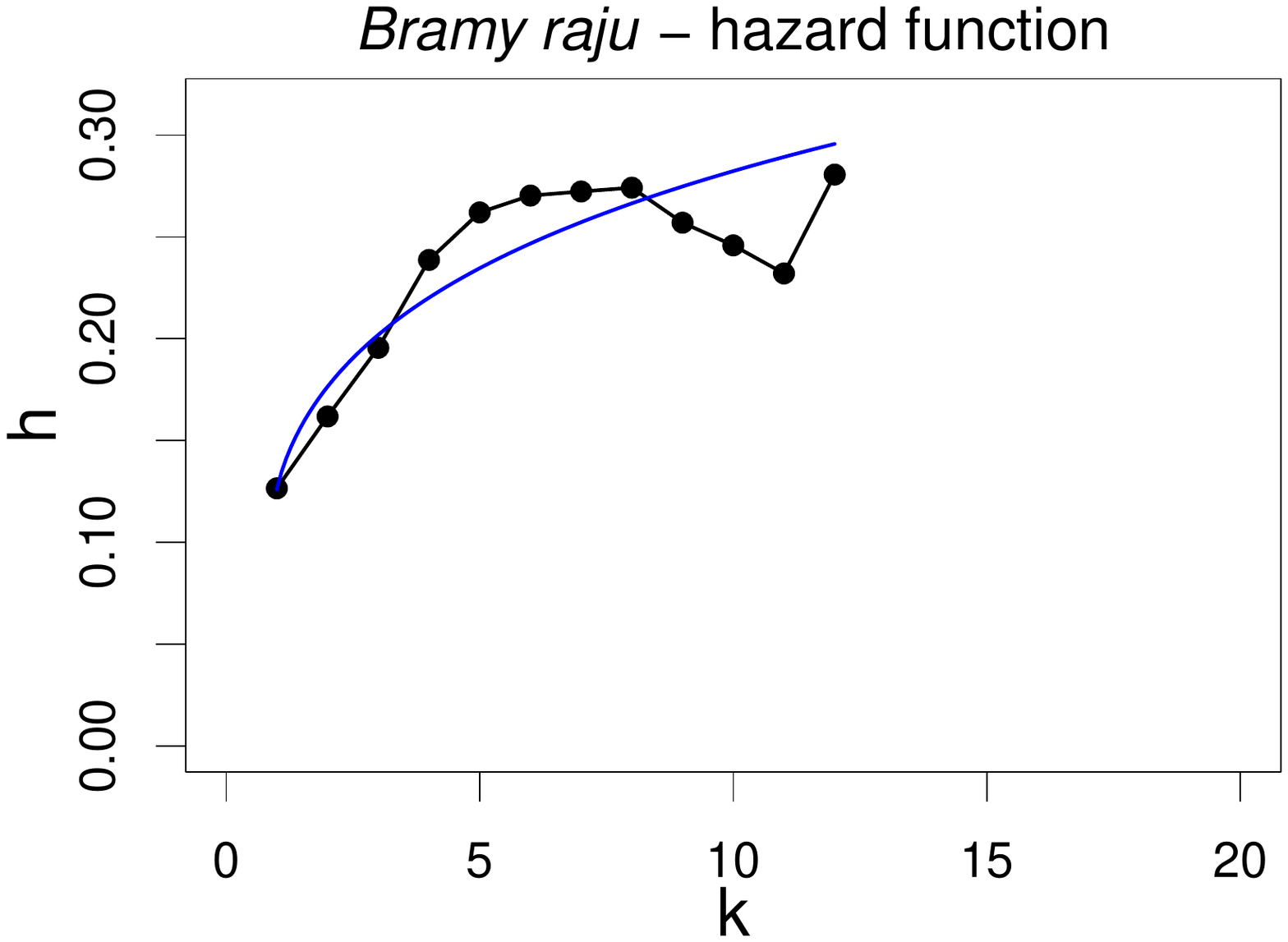}

\includegraphics[width=0.35\linewidth]{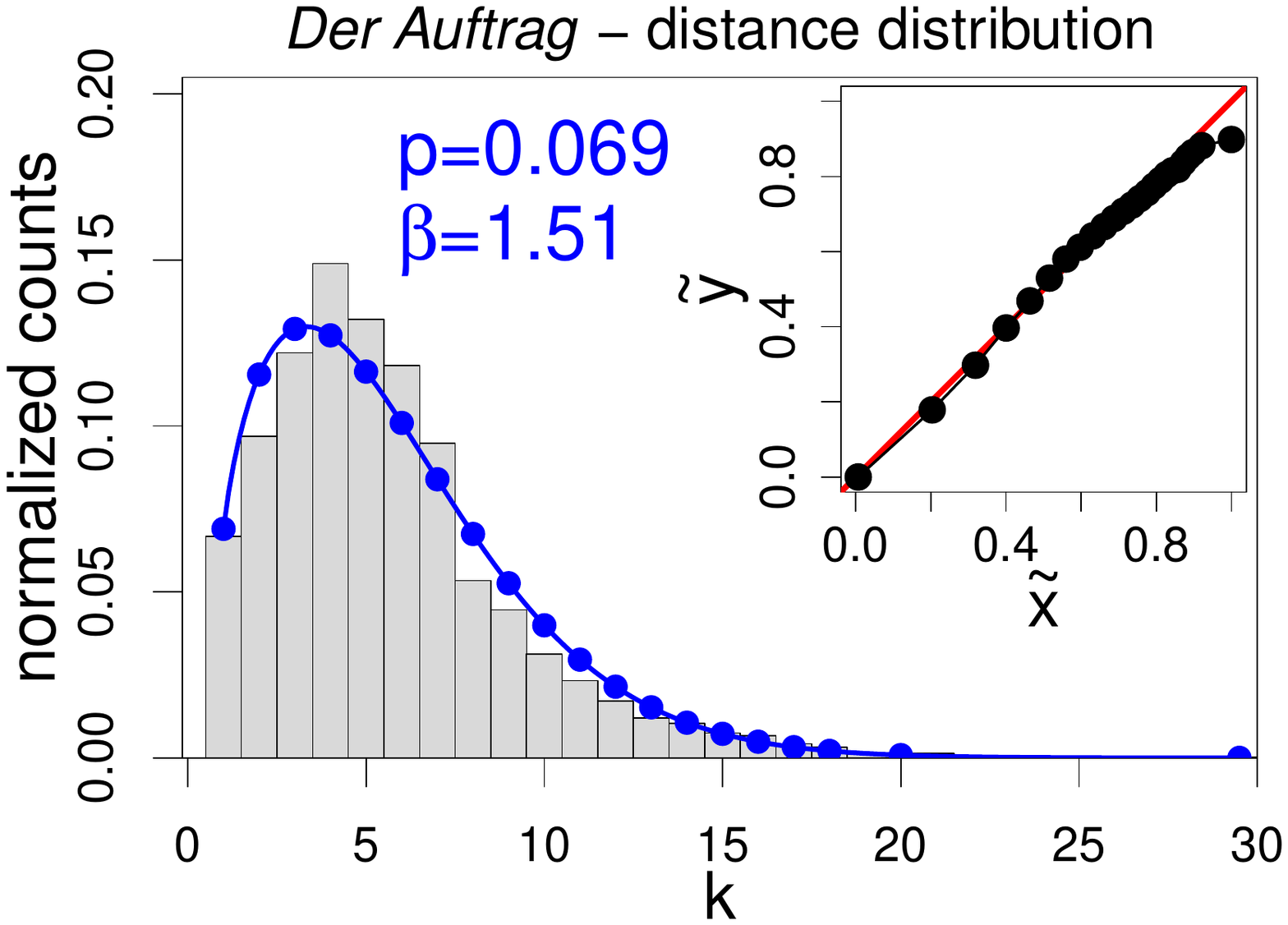}
\qquad
\includegraphics[width=0.35\linewidth]{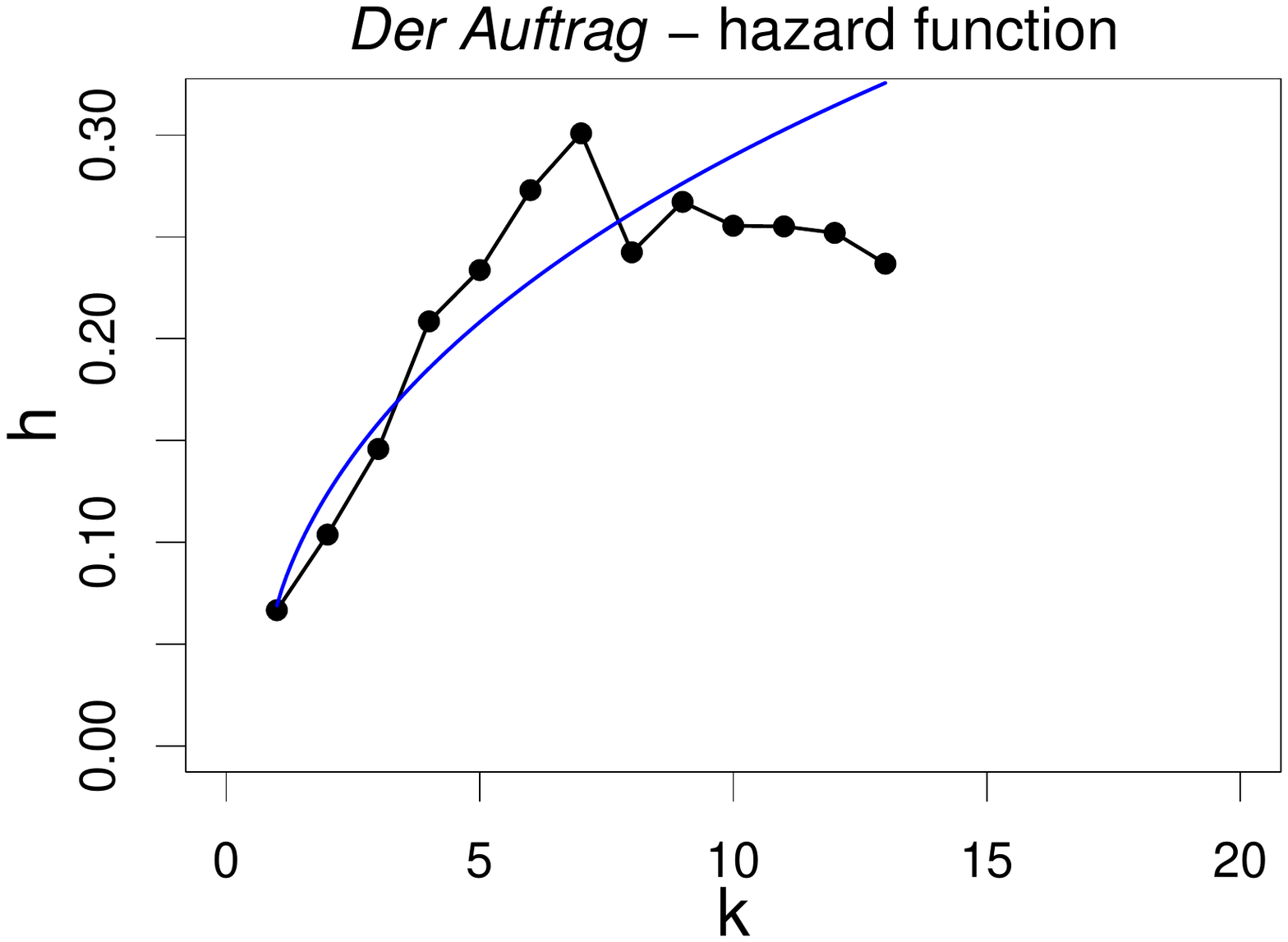}

\includegraphics[width=0.35\linewidth]{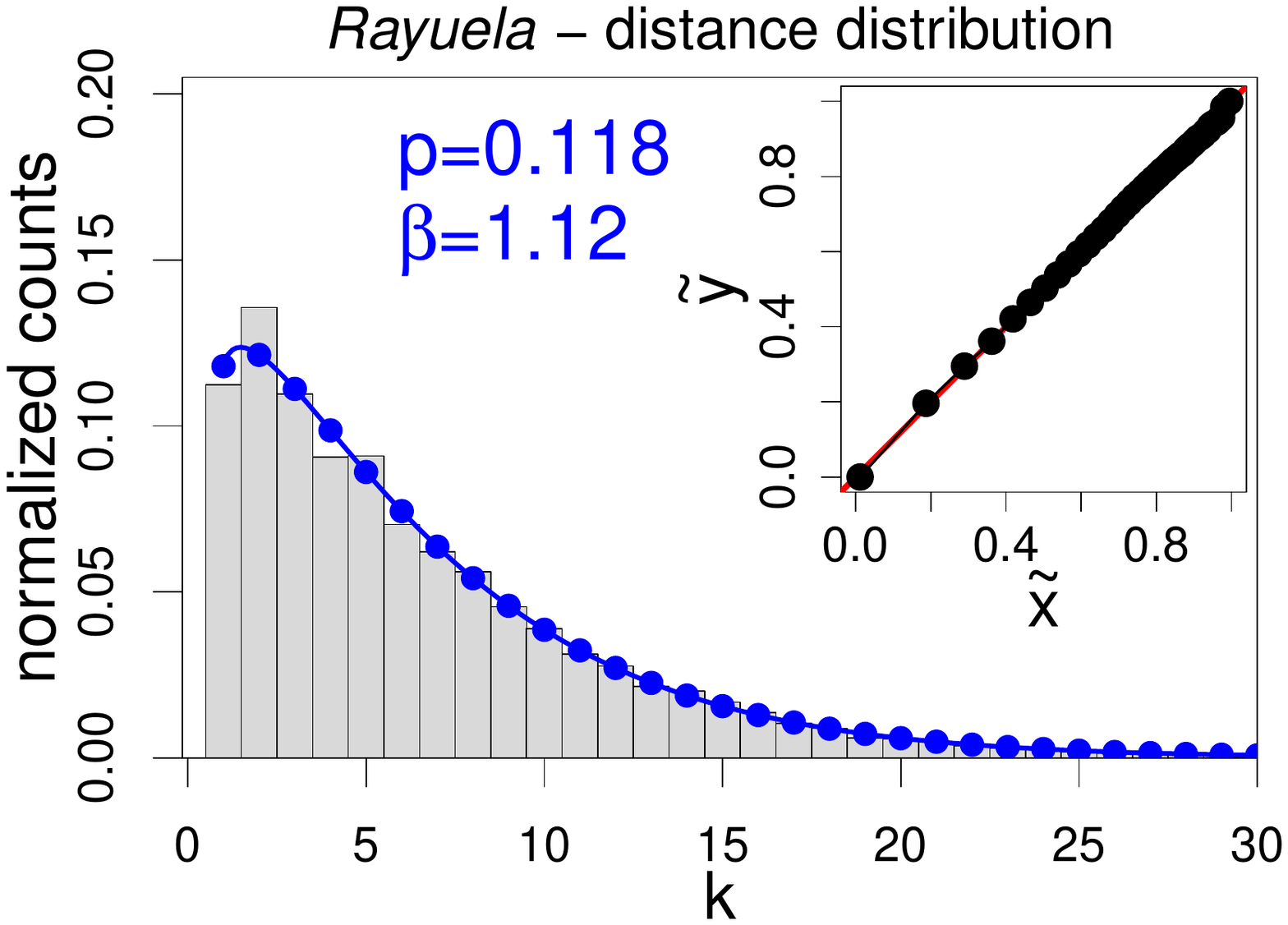}
\qquad
\includegraphics[width=0.35\linewidth]{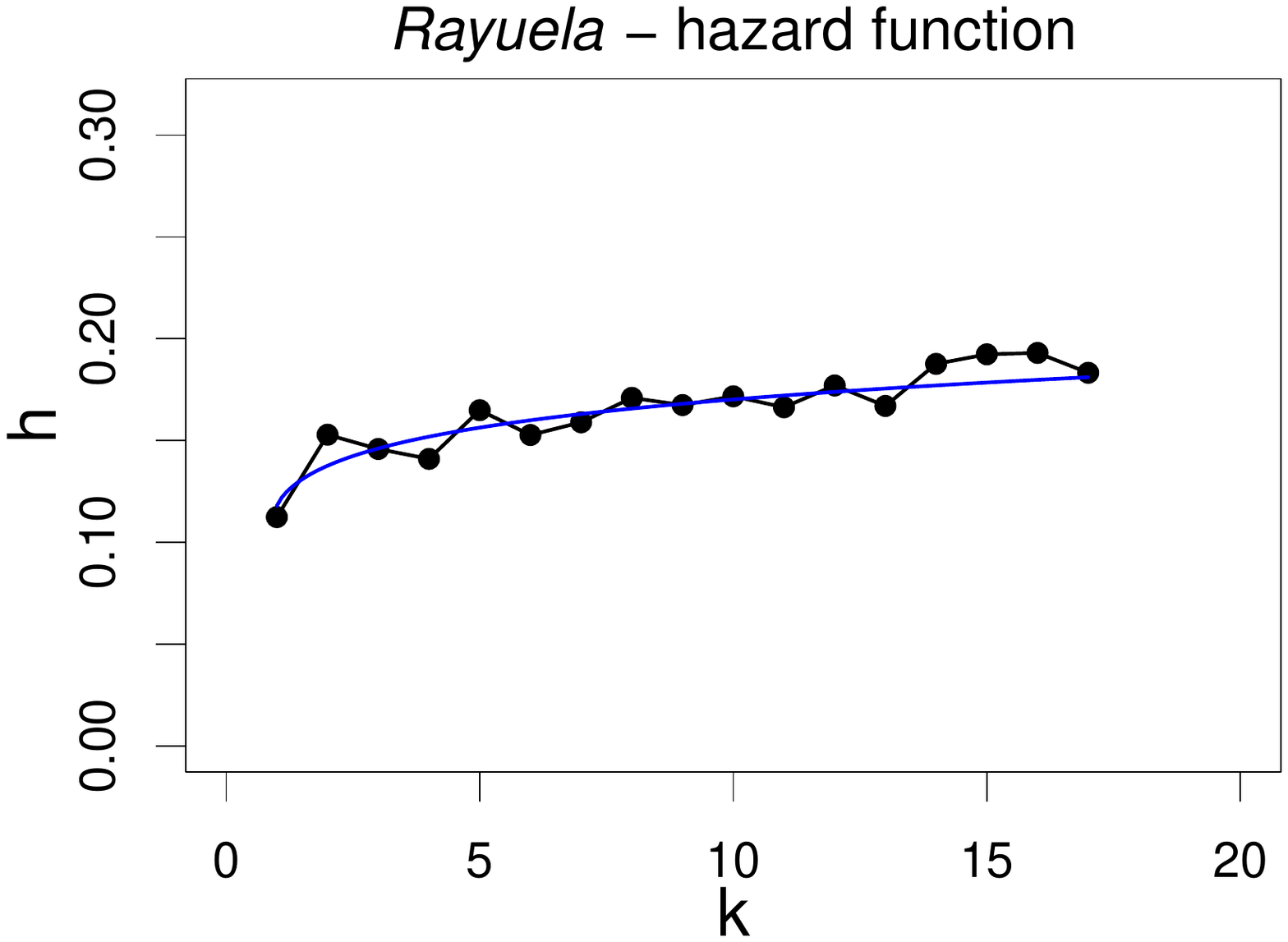}

\includegraphics[width=0.35\linewidth]{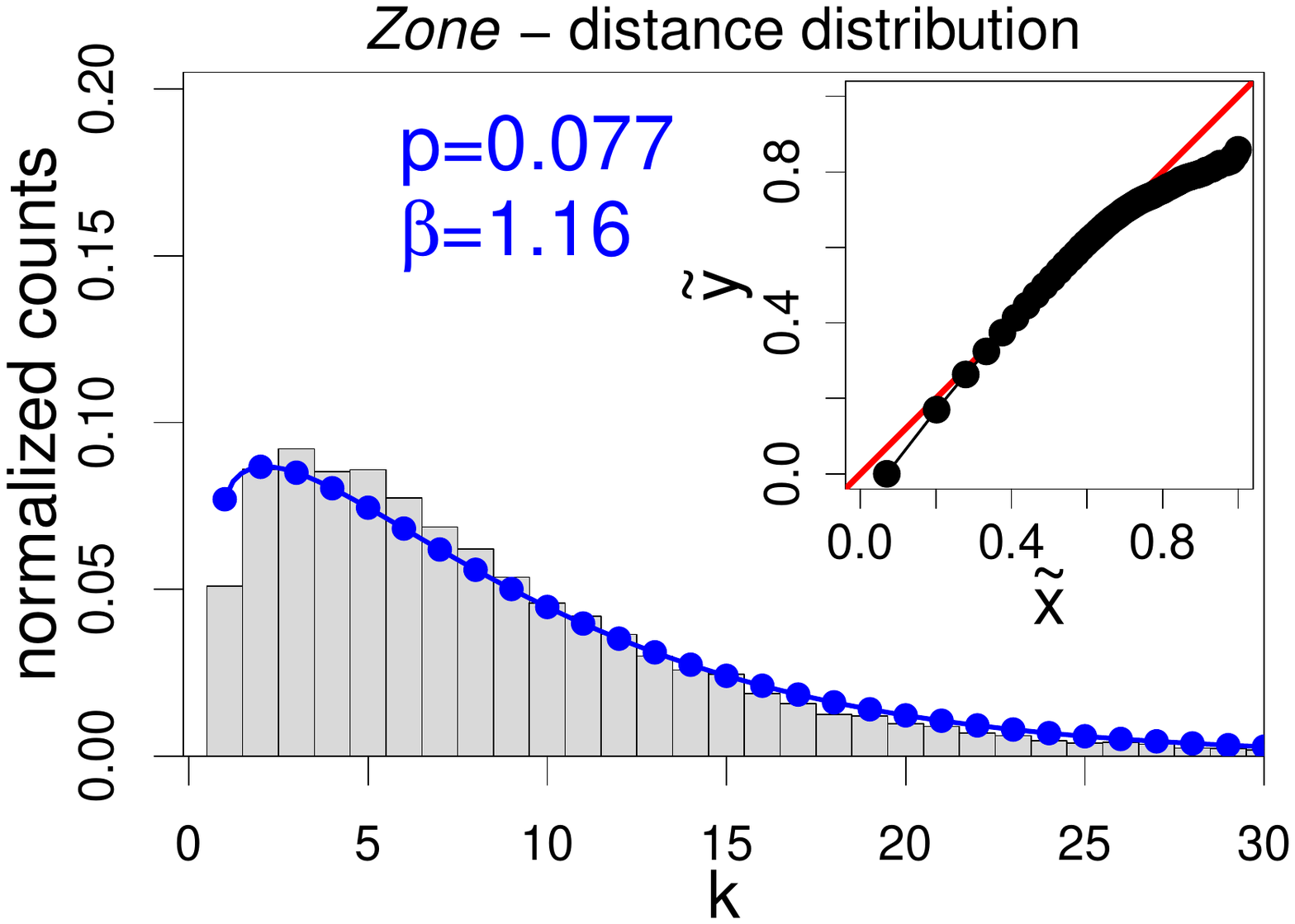}
\qquad
\includegraphics[width=0.35\linewidth]{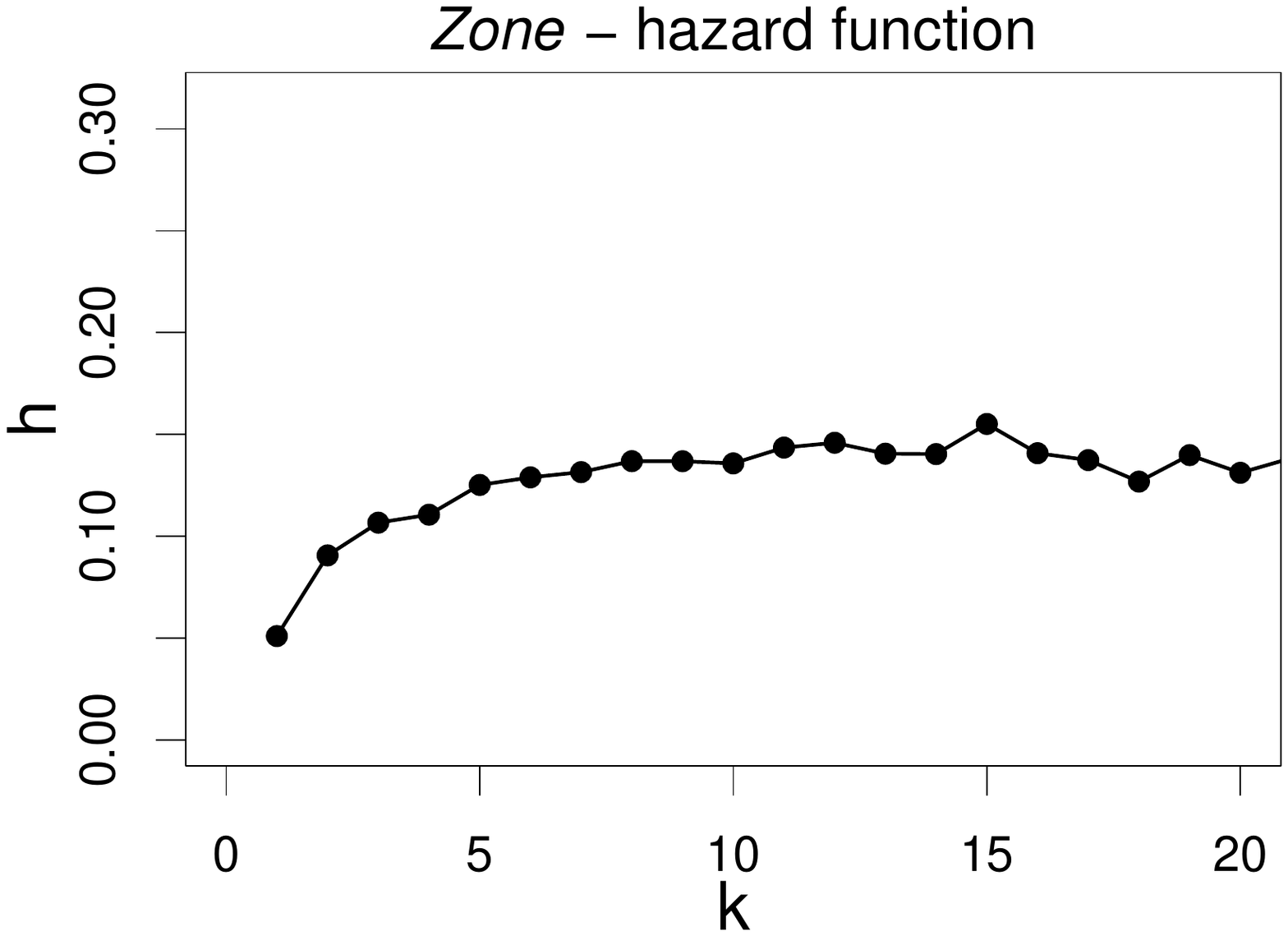}
\caption{(continued) The same characteristics for the remaining novels. For \textit{Ulysses}, the individual hazard functions of the two halves of the text are also shown.}
\end{figure*}

As the discrete Weibull distribution can correctly describe the distances between punctuation marks, even if in some texts certain characteristics of punctuation are unusual, the underlying mechanisms can be considered fairly robust~\cite{Stanisz2023,Stanisz2024}. It is possible, however, to write a text in such a way that the distribution of the inter-punctuation distances deviates from the discrete Weibull distribution. Although in the studied set of texts, the largest deviations of this type are a consequence of the presence of a few extremely long word sequences not separated by punctuation, removing such outlying observations does not restore the full agreement between the data and the model distribution. But even when the shape of the distributions is different from the one observed in more regular texts, certain statistical properties of punctuation might remain similar. An example of such a property is the monotonicity of the hazard function $h(k)$. Typically, $h(k)$ is increasing with $k$ ($\beta > 1$), which implies that $h(k) \rightarrow 1$ when $k \rightarrow \infty$. This property expresses the intuitive fact that, if the length of an unpunctuated sequence of words increases, encountering of a punctuation mark becomes more and more likely. In 8 out of 10 texts listed in Tab.~\ref{tab:analyzed_books}, the hazard function is increasing within the range of $k$ corresponding to about 95\% observations. The two exceptions are \textit{Finnegans Wake} and \textit{Ulysses} -- their hazard functions are clearly different from both the ones describing regular texts and the ones describing the other novels in Tab.~\ref{tab:analyzed_books}. In this sense, these two works of James Joyce may be considered exceptional even among the books characterized by unconventional punctuation usage patterns. Between these two, \textit{Finnegans Wake}, however, stands out more because this characteristic is uniformly distributed throughout it, while in \textit{Ulysses}, it applies only to the second half, and this is also shown in the corresponding panel of Fig.~\ref{fig:histograms_and_h}.

\begin{figure*}[!htp]
\centering
\includegraphics[width=0.35\linewidth]{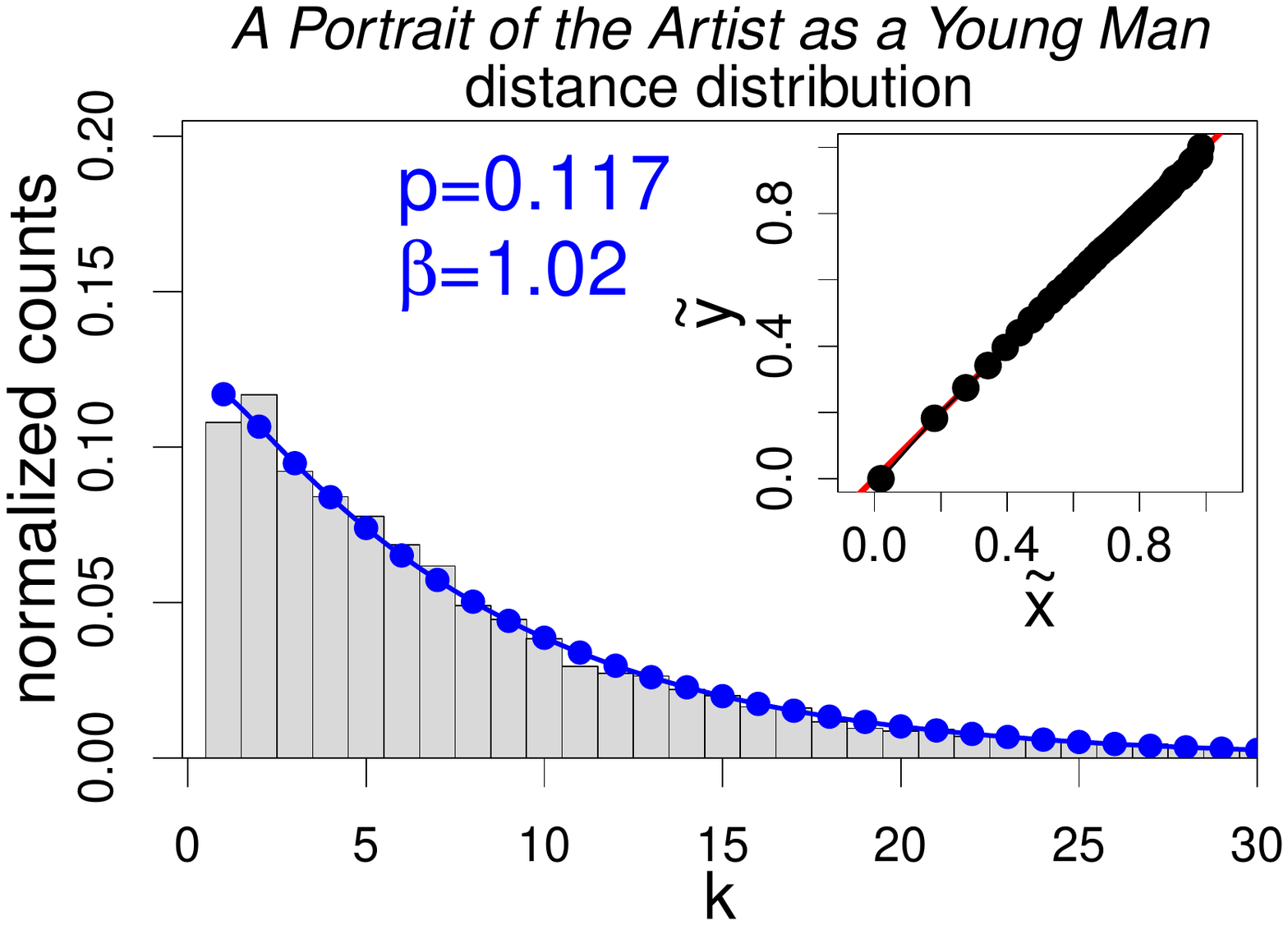}
\qquad
\includegraphics[width=0.35\linewidth]{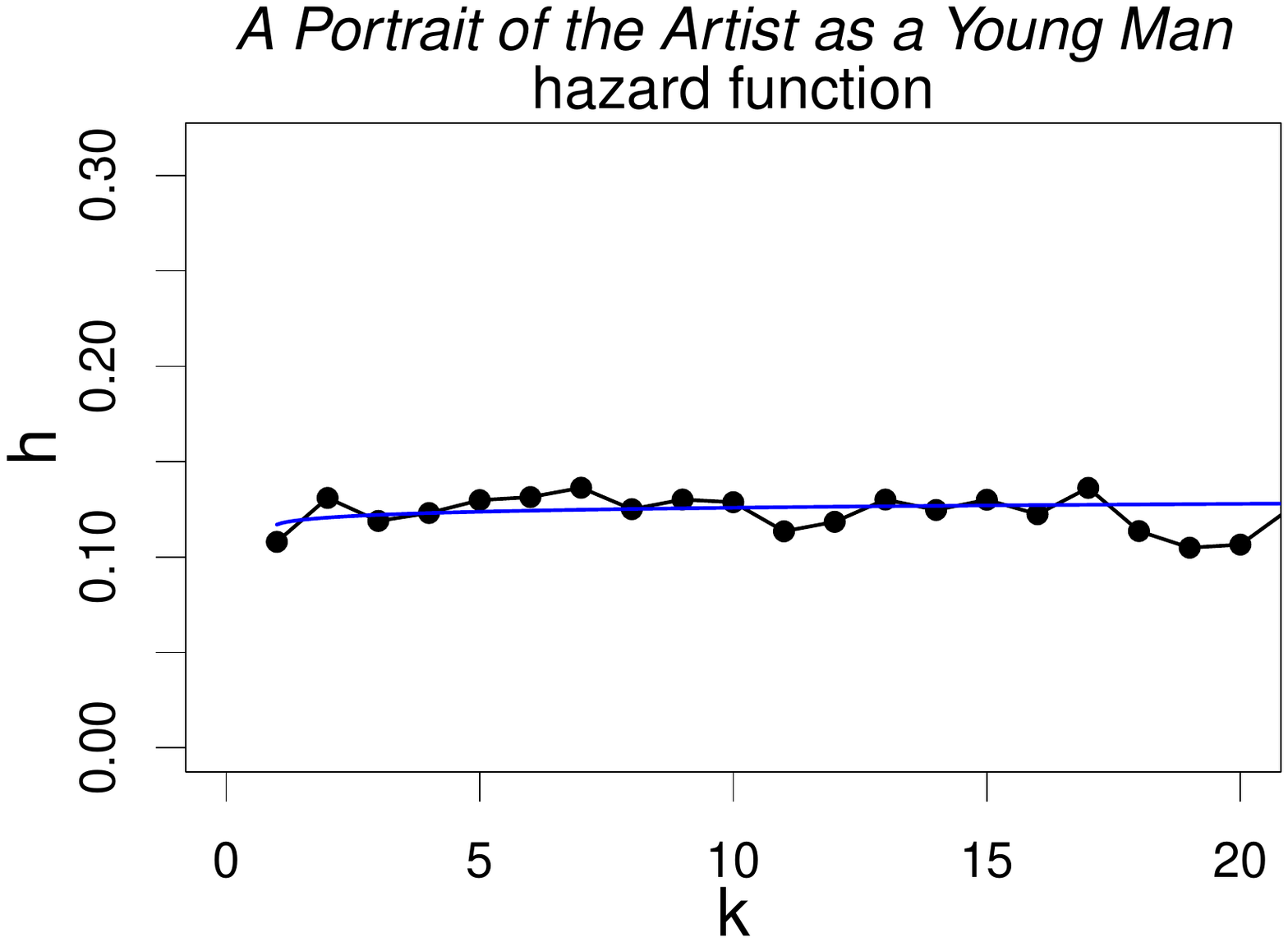}

\includegraphics[width=0.35\linewidth]{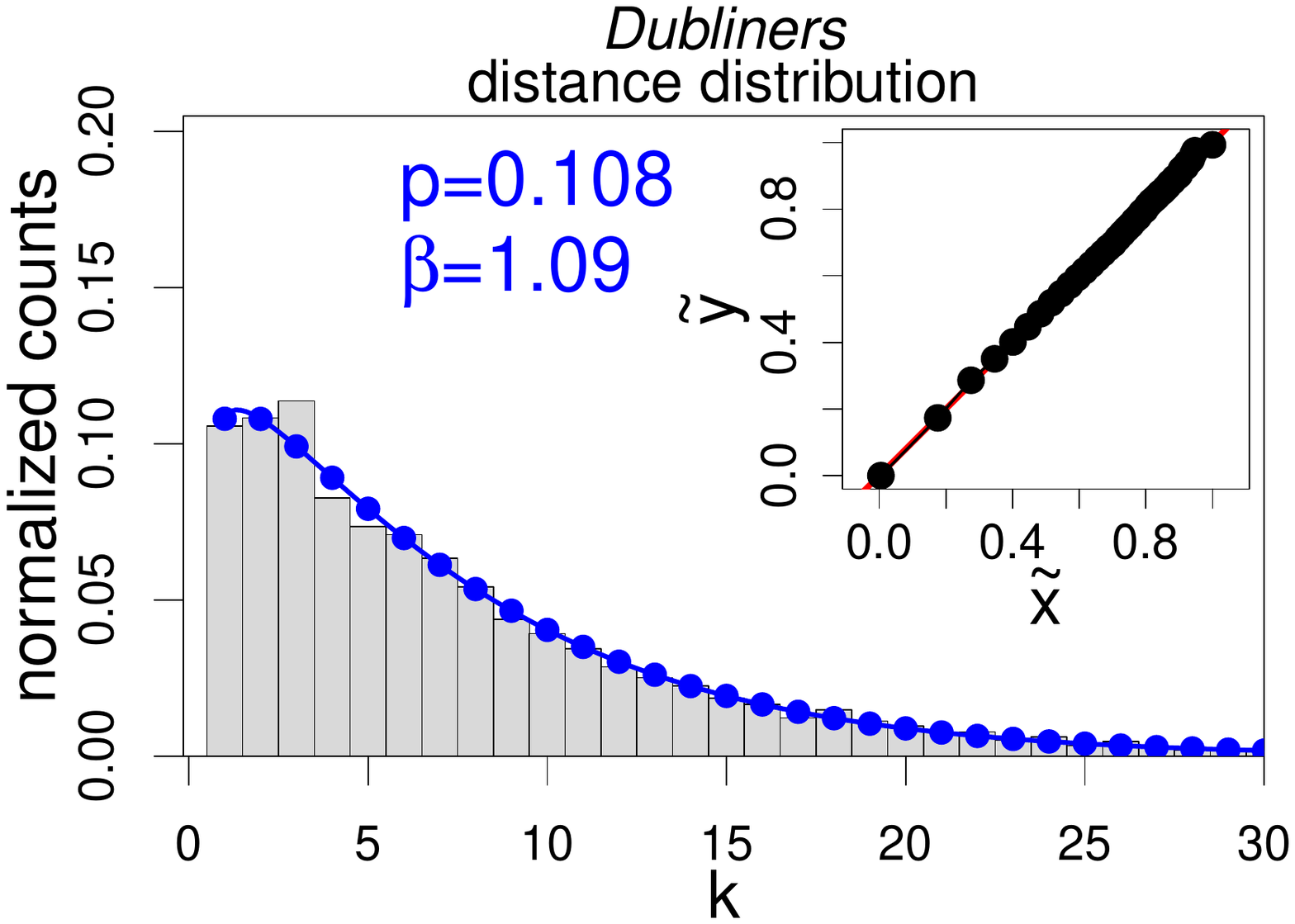}
\qquad
\includegraphics[width=0.35\linewidth]{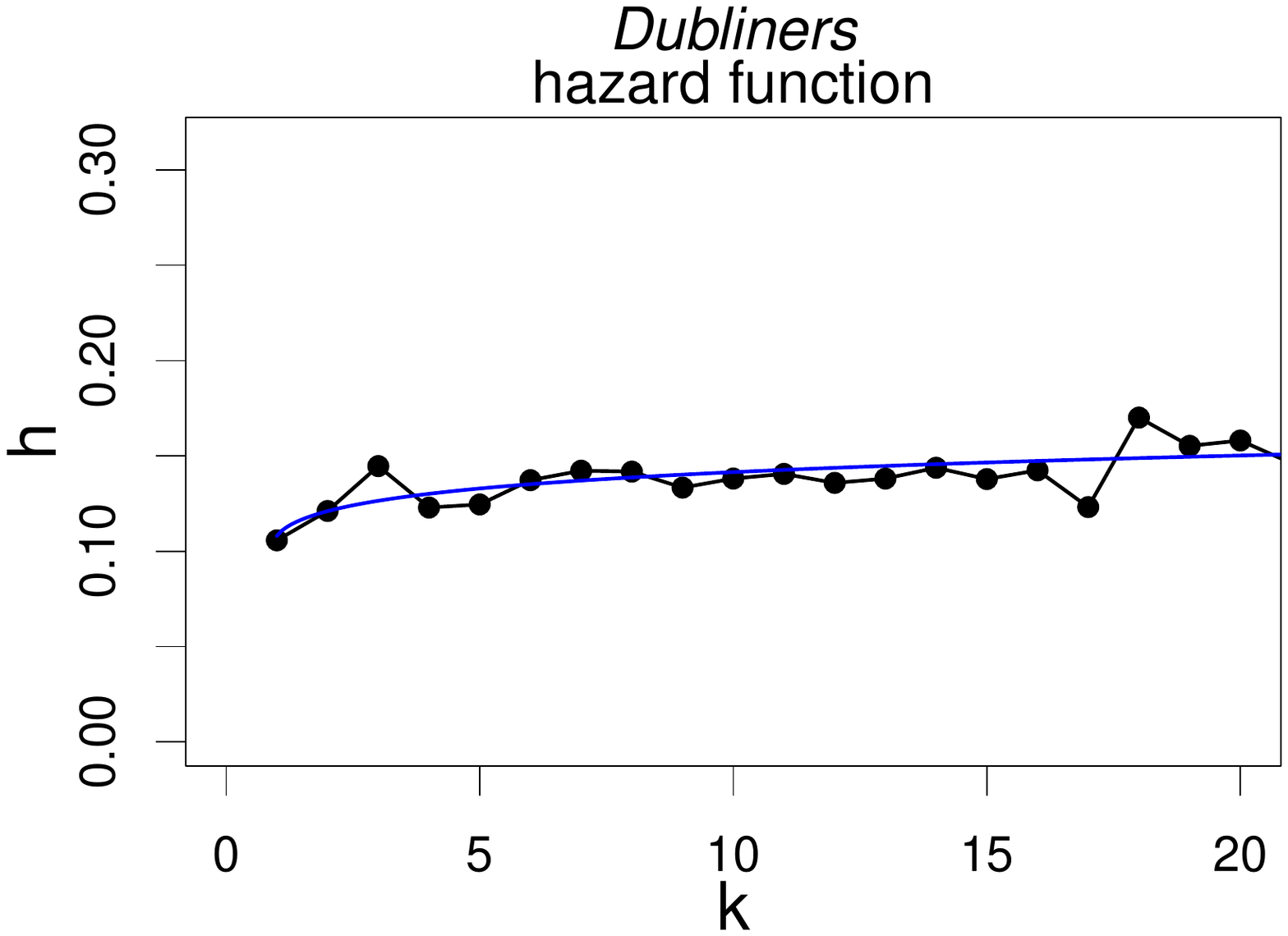}
\caption{The same characteristics as in Fig.~\ref{fig:histograms_and_h}, for two James Joyce's books: \textit{A Portrait of the Artist as a Young Man} and \textit{Dubliners} -- histograms of inter-punctuation-mark distances along with the fitted discrete Weibull distributions (left column) and the corresponding hazard functions (right column).}
\label{fig:histograms_and_h_early_Joyce}
\end{figure*}

This result provides another quantitative argument in favor of the “doubleness” of \textit{Ulysses}~\cite{McHale-1993}.
The fact that it is for the second, and thus later, half of \textit{Ulysses} that the hazard function $h(k)$ becomes decreasing (so $\beta < 1$) with increasing punctuation distance $k$ suggests a look at Joyce's earlier works. Two widely known earlier ones than Ulysses were \textit{Dubliners} and \textit{A Portrait of the Artist as a Young Man}. Characteristics analogous to those in Fig.~\ref{fig:histograms_and_h} for these two books are shown in Fig.~\ref{fig:histograms_and_h_early_Joyce}. For the chronologically first of them, \textit{Dubliners}, the hazard function still behaves typically, i.e. it is increasing, which corresponds to $\beta > 1$. For the second one, however, $\beta$ is almost equal to unity and the hazard function becomes essentially constant. Such a comparison allows us to formulate an interesting observation that this characteristics of Joyce's writing style has been progressing systematically and a clear transition to a decreasing hazard function in the use of punctuation occurred around the middle of \textit{Ulysses} and already covered the entire \textit{Finnegans Wake}. 

It is worth mentioning here that according to Joyce's division, the 18 chapters of Ulysses are divided into three parts ending with chapters 3 and 15, respectively, while the transition discussed here is observed for the partition into two parts, determined by the end of chapter 10. As it will be shown in the next section, multifractal analyzes also indicate a more complex organization of \textit{Ulysses} from roughly the middle of the book (see Fig.~\ref{fig:MFDFA_allpunctuation}).

\section{Multiscaling}

The formalism presented above refers solely to the distribution of the distances between consecutive punctuation marks; it does not specify, for example, the long-range dependencies between the distances present in different parts of a text. Investigation of the presence and the character of such dependencies can be carried out by considering time series of the distances between consecutive punctuation marks in each novel. The potential existence of the long-range dependencies can be a manifestation of the complexity of the underlying processes. A particularly important concept in that context is multifractality~\cite{JaffardS-2019a}, i.e., the presence of multiple, interwoven scaling regimes (multiscaling), which is often associated with complexity~\cite{KwapienJ-2012a}. Multifractality can be identified by means of the multifractal detrended fluctuation analysis (MFDFA)~\cite{Kantelhardt2002}, a method that is considered particularly reliable~\cite{OswiecimkaP-2006a}. MFDFA is a multiscale generalization of the detrended fluctuation analysis (DFA)~\cite{Peng1994,Kantelhardt2001} designed to estimate the Hurst exponent of a time series. Hence, MFDFA allows one to both quantify multiscaling and estimate the Hurst exponent.

MFDFA consists of the following steps. First, from the time series $x(i)$ ($i = 1, 2, ..., N$) one determines its \textit{profile} $y(i) = \sum_{k=1}^{i} x(k)$. The profile is then partitioned into disjoint segments of length $s$; one partition starts at the beginning of the time series and the other starts at the end and goes backwards -- this gives $2M_s$ segments in total. Next, the variance $\sigma^2(\nu,s)$ is computed for each segment $\nu = 1, 2, ..., 2M_s$:
\begin{equation}
\sigma^2(\nu,s) = \frac{1}{s} \sum_{k=1}^{s} \bigg( y \left( (\nu-1)s + k\right) - P_{\nu}(k) \bigg),
\end{equation}
where $P_{\nu}(k)$ is a detrending polynomial (usually of a small degree) fitted to a given segment $\nu$. Then, the $q$th-order fluctuation function is determined:
\begin{equation}
F_q(s) = \left( \frac{1}{2M_s} \sum_{\nu = 1}^{2M_s} \left( \sigma^2(\nu,s) \right)^{q/2} \right)^{1/q}
\end{equation}
for $q\neq 0$ and
\begin{equation}
F_0(s) = \frac{1}{2M_s} \sum_{\nu=1}^{2M_s} \ln \sigma^2(\nu,s)
\end{equation}
for $q=0$. The computation of $\sigma^2(\nu,s)$ and $F_q(s)$ is repeated for a range of values of $s$ and scaling behavior of $F_q$ is investigated. Observing a power-law relationship of the form
\begin{equation}
F_q(s) \propto s^{h(q)}
\end{equation}
allows one to identify the fractal properties of the studied time series: if $h(q)$ is independent of $q$, then the time series is monofractal, while an explicit dependence on $q$ indicates multifractality. $h(q)$ is called the generalized Hurst exponent, a it equals the Hurst exponent $H$ for $q=2$. The relationship between $q$ and $h$ allows for determining the Hölder (singularity) exponents:
\begin{equation}
\alpha = h(q) + q\frac{dh}{dq}
\end{equation}
and the singularity spectrum~\cite{HalseyTC-1986a}:
\begin{equation}
f(\alpha) = q \left( \alpha - h(q) \right) + 1.
\end{equation}
The function $f(\alpha)$ can be interpreted as the fractal dimension of a set of points characterized by the singularity exponent $\alpha$. Shape of the singularity spectrum reflects multiscaling properties of the time series: $f(\alpha)$ collapses to a single point for a monofractal time series and it has the shape resembling a parabola opening down for a multifractal one. The width $\Delta\alpha = \alpha_{\rm max}-\alpha_{\rm min}$ of the singularity spectrum quantifies how rich is multifractality and it is therefore often considered as a measure of complexity~\cite{KwapienJ-2012a}.

\begin{figure*}[p]
\centering
\captionsetup[subfigure]{justification=centering}
\subfloat[\textit{Rayuela}]{\includegraphics[width=0.75\linewidth]{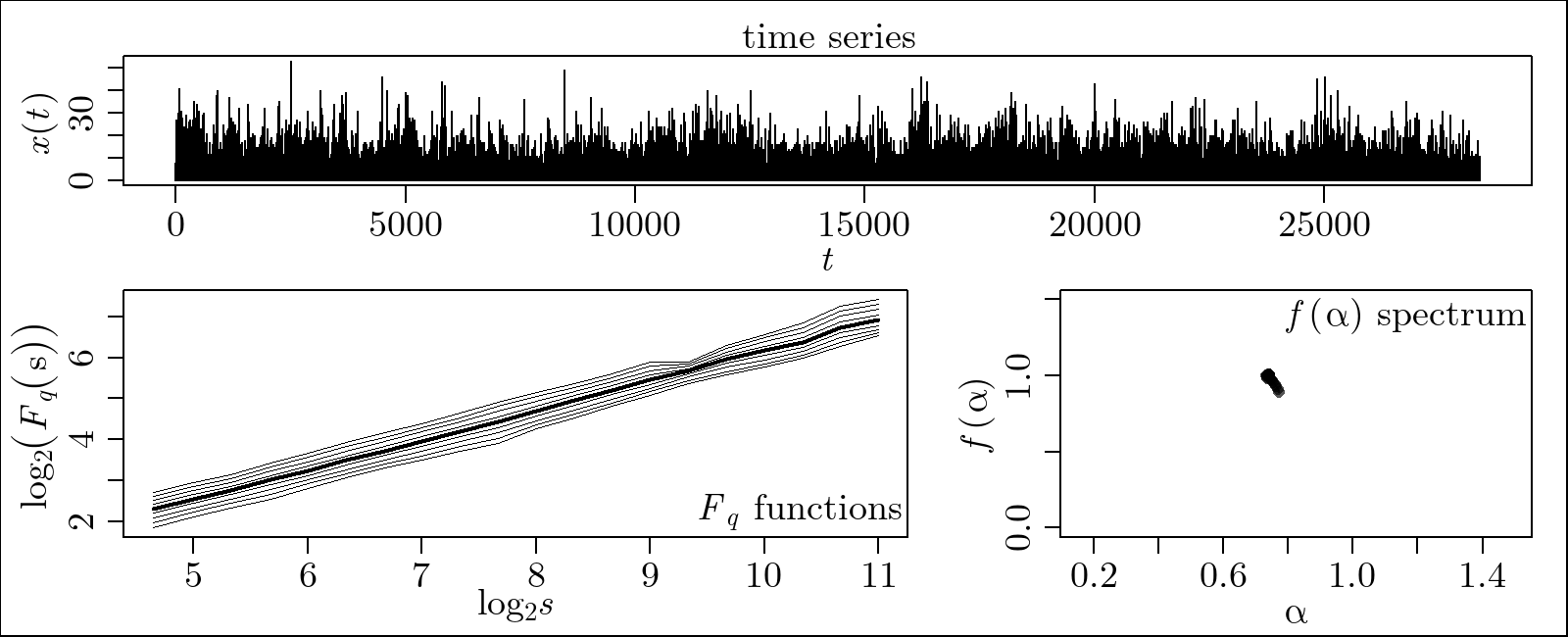}}
\vspace{2em}
\subfloat[\textit{Finnegans Wake}]{\includegraphics[width=0.75\linewidth]{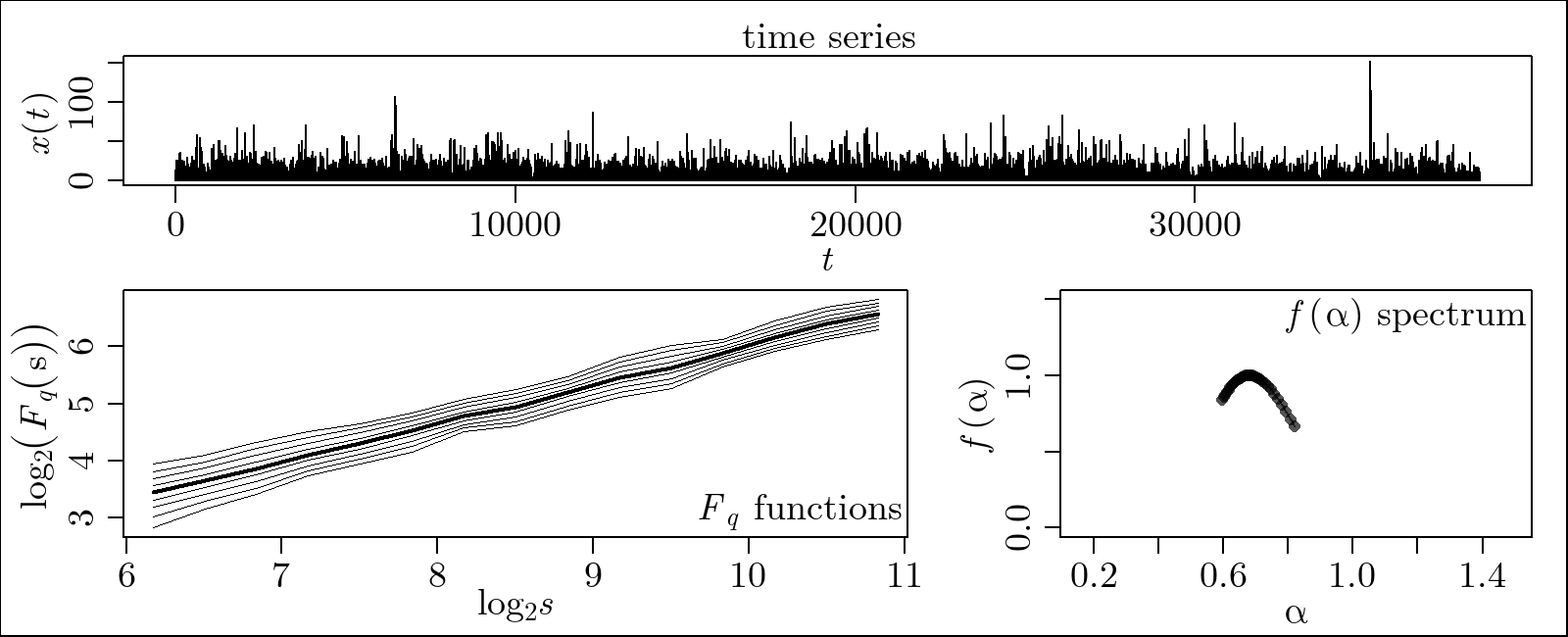}}
\vspace{2em}
\subfloat[\textit{\textit{Ulysses}}]{\includegraphics[width=0.75\linewidth]{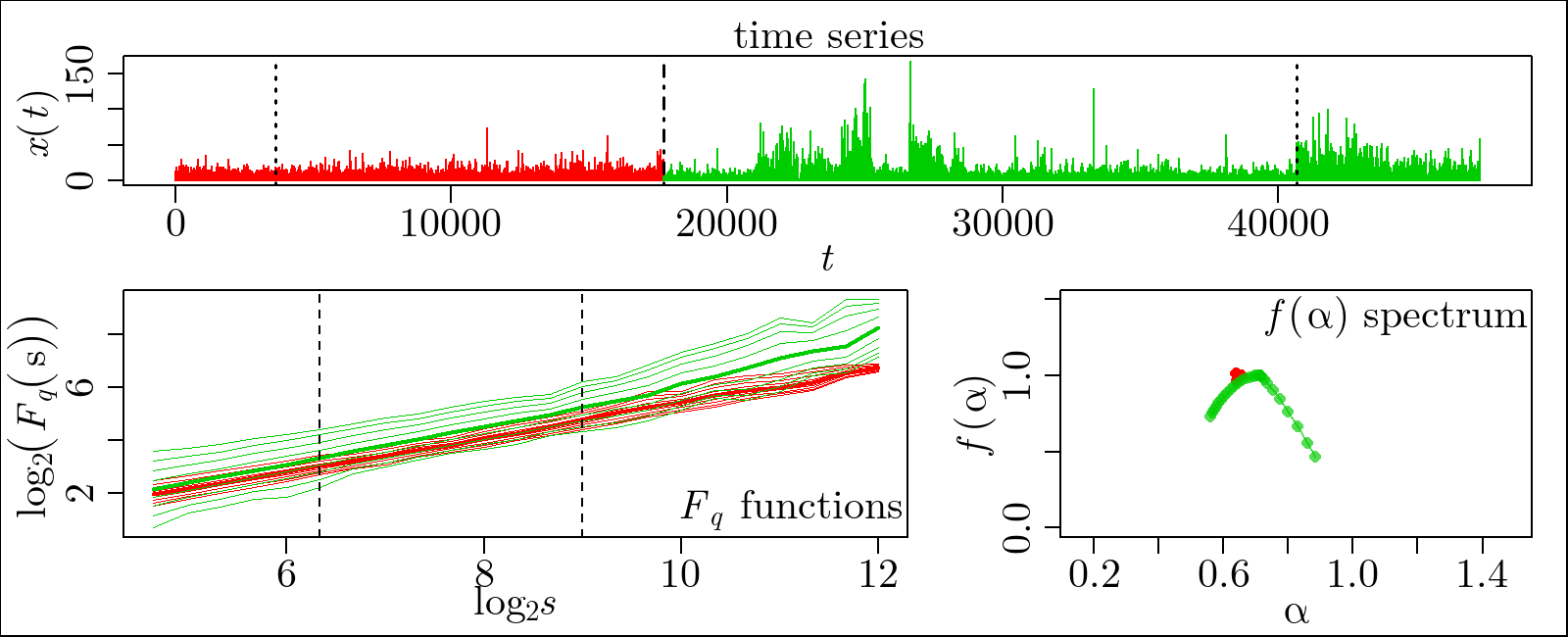}}
\vspace{1em}
\caption{MFDFA applied to time series of distances between consecutive punctuation marks for (a) \textit{Rayuela}, (b) \textit{Finnegans Wake}, and (c) \textit{Ulysses}. For each book, the original time series $x(t)$ (top), the $q$th-order fluctuation functions $F_q(s)$ (bottom left), and the singularity spectrum $f(\alpha)$ (bottom right) are shown. The fluctuation functions for $q=0$ are distinguished by bold lines. \textit{Ulysses} has been divided in two parts: the first contains chapters 1-10 (plotted in red), the second starts with chapter 11 (plotted in green). As these two parts differ qualitatively in terms of the studied characteristics, they have been analyzed separately; the point separating them (the end of chapter 10) is marked by a vertical dotted-dashed line in the relevant $x(t)$ plot. The same plot shows the end of chapter 3 and the end of chapter 15 (dotted lines), which constitute the partition into 3 parts specified in the book itself (not considered in the analysis; parts 1 and 3 in separation are too short in for such an analysis of statistical character). In the $F_q(s)$ plot, vertical dashed lines mark the range of scaling used in the computation of the $f(\alpha)$ spectrum.}
\label{fig:MFDFA_allpunctuation}
\end{figure*}

\begin{figure*}[p]
\centering
\captionsetup[subfigure]{justification=centering}
\subfloat[\textit{Rayuela}]{\includegraphics[width=0.75\linewidth]{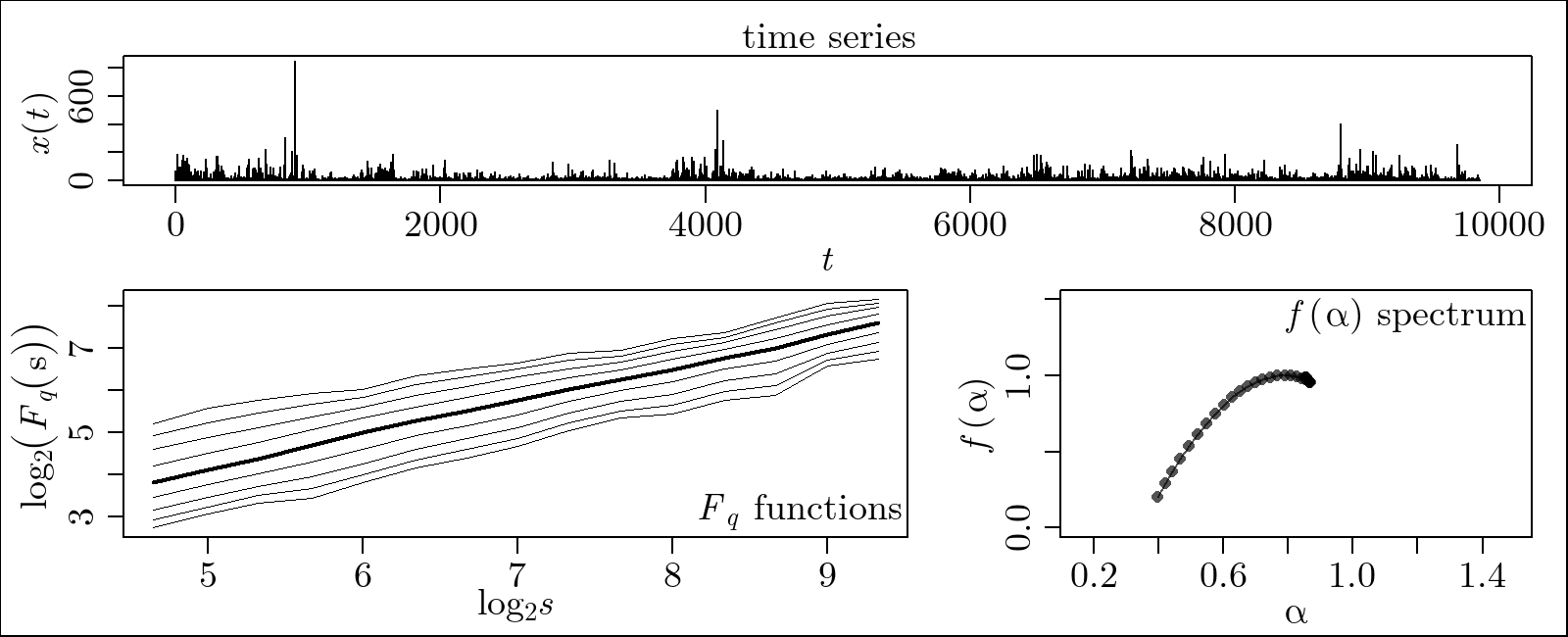}}
\vspace{2em}
\subfloat[\textit{Finnegans Wake}]{\includegraphics[width=0.75\linewidth]{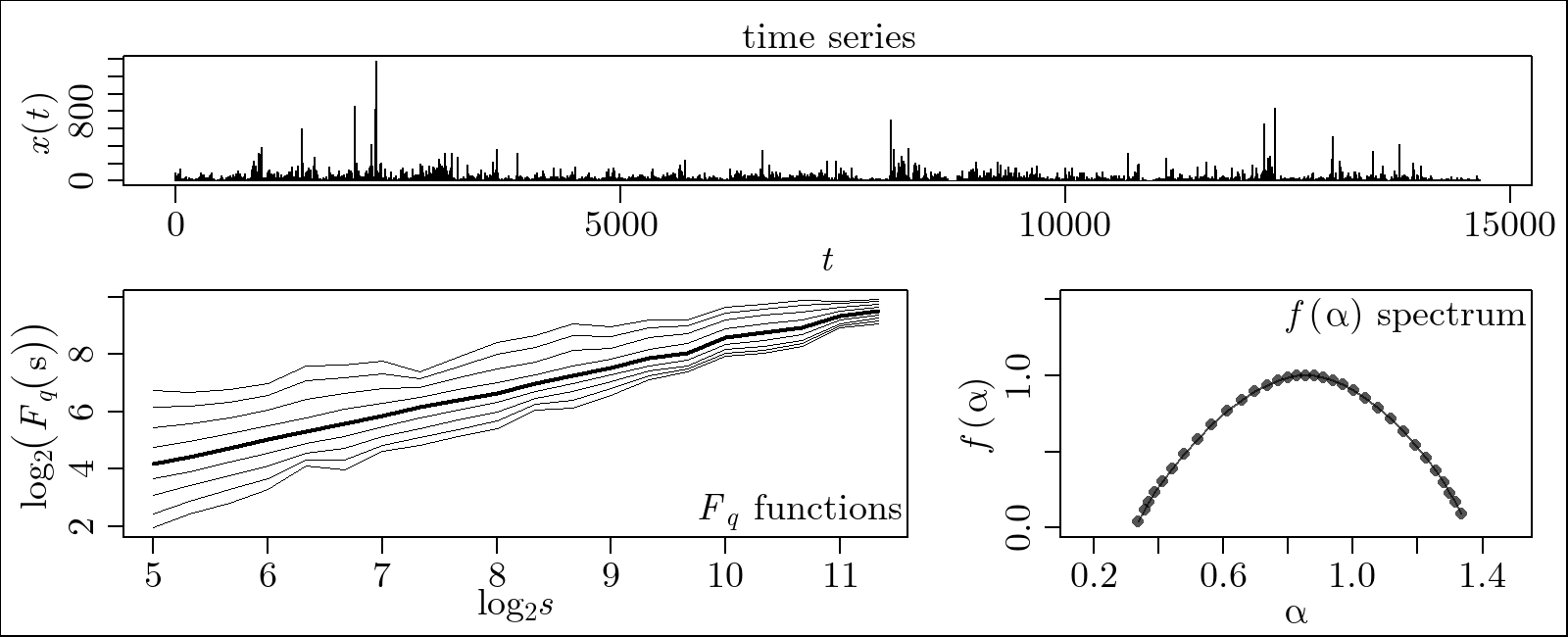}}
\vspace{2em}
\subfloat[\textit{\textit{Ulysses}}]{\includegraphics[width=0.75\linewidth]{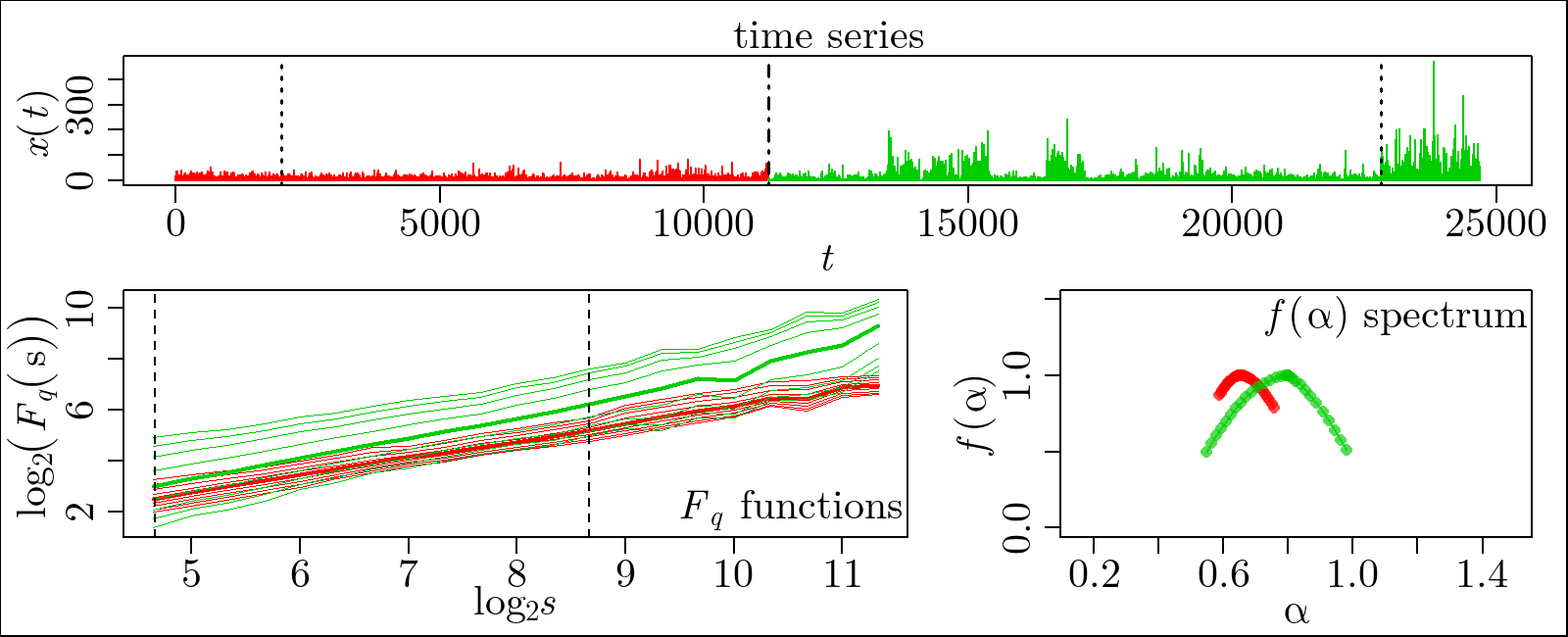}}
\vspace{1em}
\caption{MFDFA applied to time series of sentence lengths for (a) \textit{Rayuela}, (b) \textit{Finnegans Wake}, and (c) \textit{Ulysses}. The same characteristics as in Fig.~\ref{fig:MFDFA_allpunctuation} are shown.}
\label{fig:MFDFA_sentences}
\end{figure*}

Figs.~\ref{fig:MFDFA_allpunctuation} and \ref{fig:MFDFA_sentences} show the results of MFDFA applied to the time series of distances between consecutive punctuation marks and the time series of sentence lengths for \textit{Rayuela}, \textit{Finnegans Wake}, and \textit{Ulysses}, respectively. It can be observed that, if complete punctuation is taken into account, the singularity spectra $f(\alpha)$ are relatively narrow if compared to the ones corresponding to the sentence-ending punctuation marks. For \textit{Rayuela} and the first part of \textit{Ulysses}, the width for their $f(\alpha)$ is seen to be indicating their essentially monofractal character, i.e., they do not develop significant multiscaling (Fig.~\ref{fig:MFDFA_allpunctuation}(a)(c)). This is, however, not the case if \textit{Finnegans Wake} is considered: the width of $f(\alpha)$ reveals that moderately rich multifractality can be detected here (Fig.~\ref{fig:MFDFA_allpunctuation}(b)). The second part of \textit{Ulysses} seems a similar in that regard (Fig.~\ref{fig:MFDFA_allpunctuation}(c)); however, this particular result should be interpreted with caution, as the identified range of fluctuation functions' power-law behavior is relatively narrow.

In contrast, the sentence lengths, which reveal more freedom as regards the constraints imposed by the discrete Weibull distribution~\cite{Stanisz2023}, organize themselves into multifractal structures evidenced by the $f(\alpha)$ spectra of considerable width (Fig.~\ref{fig:MFDFA_sentences}). It has already been shown~\cite{Drozdz2016} that such structures are often present in texts representing the stream-of-consciousness literary style. In \textit{Ulysses} (Fig.~\ref{fig:MFDFA_sentences}(c)), multifractality pertains mainly to the second half of the book (chapters 11-18), because $f(\alpha)$ corresponding to the first half (chapters 1-10) is relatively narrow and, thus, similar to the spectra observed in the texts with a more regular narrative style. This allows one to consider \textit{Ulysses} as a work composed of two structurally different parts with the second part characterized by richer multifractal characteristics. A unique feature of \textit{Finnegans Wake}, distinguishing it even among the texts characterized by highly unusual style, is the fact that its singularity spectrum has a high level of symmetry (Fig.~\ref{fig:MFDFA_sentences}(b)). This implies the presence of a well-organized, complex hierarchy since such symmetry of $f(\alpha)$ is characteristic for model objects exhibiting perfect mathematical multifractality and it is not so often observed in real-world systems. On the other hand, the left-hand side asymmetry seen in $f(\alpha)$ for \textit{Rayuela} is much more typical for such systems; it occurs in situations where a principal carrier of multiscaling are large fluctuations while small ones are characterized by monoscaling. However, the opposite asymmetry that is seen for the first half of \textit{Ulysses} is somewhat less common and pertains to a situation where, predominantly, small fluctuations are multiscaling~\cite{DrozdzS-2015a}.

The above results on multiscaling have been confronted against their surrogate counterparts which are shown in Fig.~\ref{fig:MFDFA_surrogates} in Appendix \ref{appendix_MFDFA_surrogates}. The two types of surrogates most commonly used in this context include the Fourier-phase-randomized series and, the second one, the series obtained from the original series by a random shuffling. The first of them destroys the nonlinear correlations in the series but preserves the linear ones. In the present case the MFDFA procedure applied to such surrogates leads to a singularity spectrum which gets shrunk to essentially a point centered at around the maximum of the original one. The second way of generating surrogates, the one based on the entire randomization, destroys all the correlations and for sufficiently long series is expected to result in a monofractal spectrum located at $\alpha \approx 0.5$ (or a bifractal for fluctuations from the L\'evy stable regime), which manifests itself in $q$-independence of the fluctuation functions~\cite{Kwapien2023}. Convergence to such a result with increasing the time series length is typically very slow~\cite{Drozdz2009}. From the perspective of the fluctuation functions, the scenario of approaching this limit involves a characteristic ``whisk-shape'' seen at the smaller scales $s$. For the frequent case of shorter time series of the order of a few thousand data points, as in the present study of literary works, only this region of scales is accessible in MFDFA. As one consequence, even though the family of $F_q(s)$ does not develop a real multifractal scaling for randomized data, such an apparent multifractality is often mistakenly interpreted as a genuine one. Such a ``whisk-shape'' -- and therefore the remaining width of the singularity spectrum -- as a finite-size effect for shuffled surrogates is also seen here for those series that originally display multiscaling. Hence, true multiscaling pertains only to the original series, not to the randomized ones.

\section{Conclusions}

In this study, statistical characteristics of punctuation in a set of literary works with unusual use of punctuation have been considered. The analysis focused on the time series of inter-punctuation-mark distances and of sentence lengths, expressed in the number of words. These time series have been modeled with the discrete Weibull distribution and in terms of the related hazard functions, describing how likely it is to encounter a punctuation mark (any punctuation mark or a sentence-ending mark) depending on the length of the preceding unpunctuated word sequence. It has been found that the discussed properties stay within the same regime that can be observed in texts for which the writing style (and, consequently, punctuation usage) is considered typical. However, there exist texts with clearly different properties. The most distinctive examples among the considered novels are \textit{Ulysses} and \textit{Finnegans Wake} by James Joyce. In these two texts (in a half of the former one and in the whole latter one), the distributions of the distances between punctuation marks can be characterized by decreasing hazard functions. This result may be viewed as a signature of an ability and preference to compose uninterrupted linguistic constructs that have a tendency to grow more the longer they have already been generated. At the same time, the works of Joyce are distinctive also in terms of certain other properties related to punctuation, like the long-range correlations arranging the sentence length variability in cascading patterns. The presence of such correlations is related to the presence of multifractal structures. While multifractality can also be observed in some other literary texts that use the stream-of-consciousness narrative technique, an exceptionally rich hierarchy of scaling, comparable to the one observed in mathematically idealized multifractal systems, is identified in \textit{Finnegans Wake}. Interestingly, \textit{Finnegans Wake} exhibits a trace of multifractality also with respect to all punctuation marks.

\section*{Data Availability Statement}

The data that support the findings of this study are available from the corresponding author upon reasonable request.

\appendix

\section{Transformation of the Weibull plot coordinates $(x,y) \longrightarrow (\widetilde{x},\widetilde{y})$ \label{appendix_Weibull_plot}}

If some empirical data come from the discrete Weibull distribution with parameters $(p,\beta)$, then a straight line with slope $\beta$ and intercept $\log\left( -\log \left( 1-p \right) \right)$ should be observed while plotting the empirical cumulative distribution function $\mathcal{F}_{\rm emp}(k)$ in the coordinates $(x,y)$, where 
\begin{align*} 
&x=\log k \\
&y=\log\left( -\log \left( 1-\mathcal{F}_{\rm emp}(k) \right) \right).
\end{align*}
A deviation of the empirical distribution from the model distribution is observed as a deviation of the former from a straight line. To obtain a rescaled plot, which fits in the square $[0,1]\!\times\![0,1]$ and has the reference line with slope 1 and intercept 0, one applies the following transformation. Let $(x_{\rm min},x_{\rm max},y_{\rm min},y_{\rm max})$ be the minimum and the maximum value of $x$ and $y$ appearing on a given Weibull plot, respectively and let $y=a+bx$ be a line representing the model Weibull distribution. With the quantities defined as follows:
\begin{align*} 
&x_{\rm plot.min} = \min \left\{ x_{\rm min}, \, \frac{y_{\rm min}-a}{b} \right\} \\
&x_{\rm plot.max} = \max \left\{ x_{\rm max}, \, \frac{y_{\rm max}-a}{b} \right\} \\
&y_{\rm plot.min} = \min \left\{ y_{\rm min}, \, a + b\,x_{\rm min} \right\} \\
&y_{\rm plot.max} = \max \left\{ y_{\rm max}, \, a + b\,x_{\rm max} \right\},
\end{align*}
the transformation from $(x,y)$ to $(\widetilde{x},\widetilde{y})$ is given by
\begin{align*} 
&\widetilde{x}=\frac{x-x_{\rm plot.min}}{x_{\rm plot.max}-x_{\rm plot.min}} \\
&\widetilde{y}=\frac{y-y_{\rm plot.min}}{y_{\rm plot.max}-y_{\rm plot.min}}.
\end{align*}

\section{MFDFA surrogates \label{appendix_MFDFA_surrogates}}
\begin{figure*}[p]

\newlength{\SingleSurrogatePlotWidth}
\setlength{\SingleSurrogatePlotWidth}{0.485\linewidth}

\centering
\includegraphics[width=\SingleSurrogatePlotWidth]{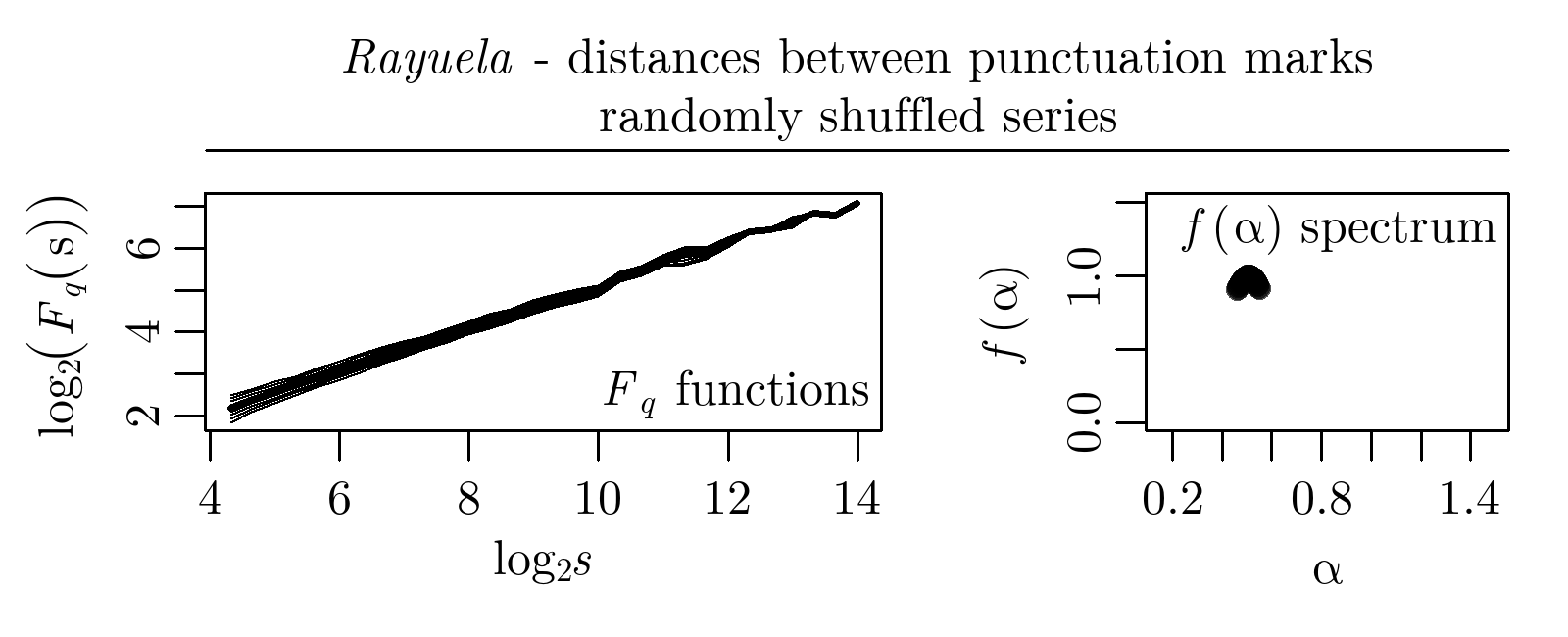}
\hfill
\includegraphics[width=\SingleSurrogatePlotWidth]{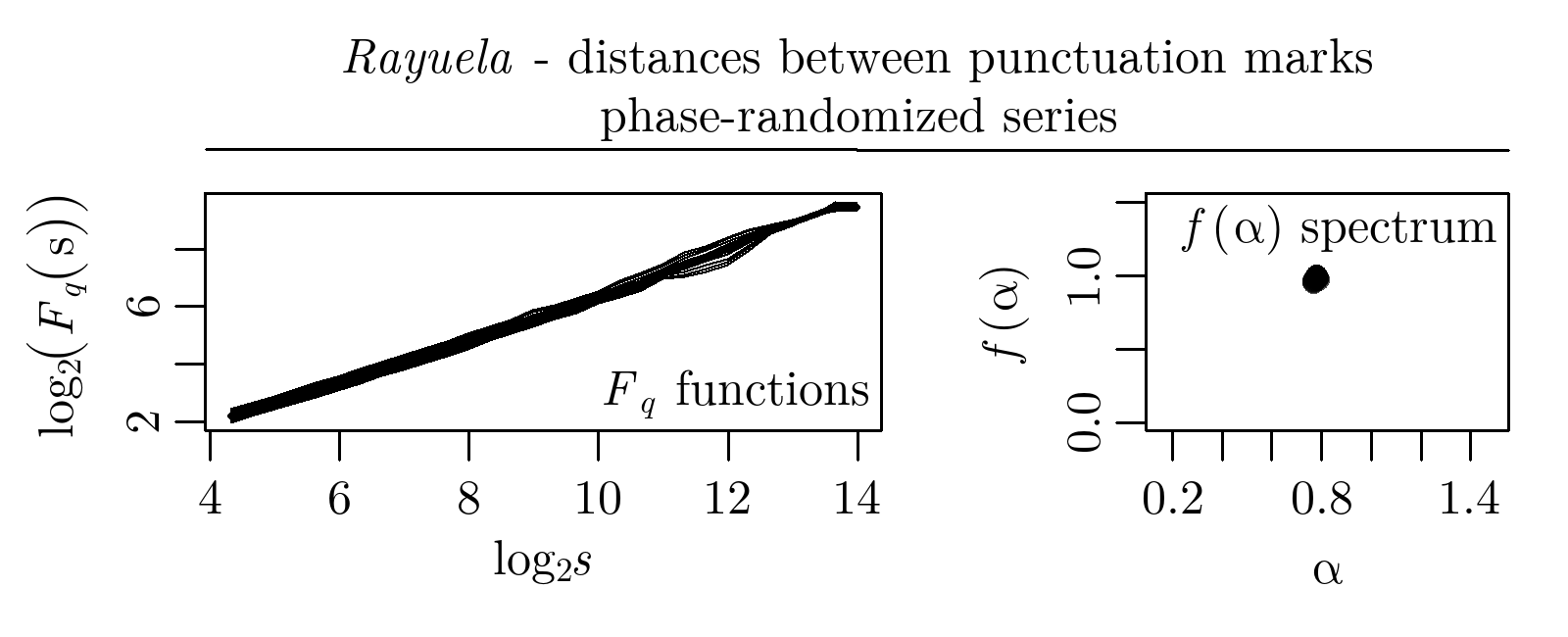}

\includegraphics[width=\SingleSurrogatePlotWidth]{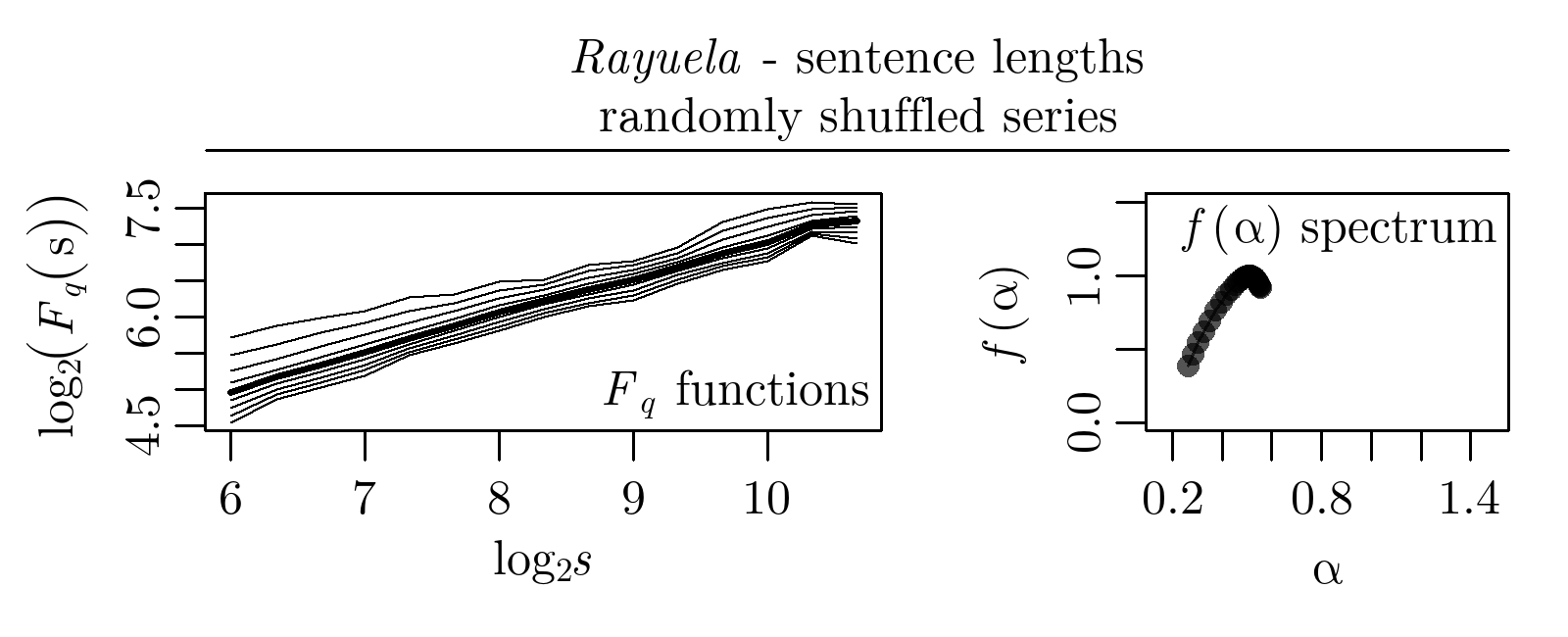}
\hfill
\includegraphics[width=\SingleSurrogatePlotWidth]{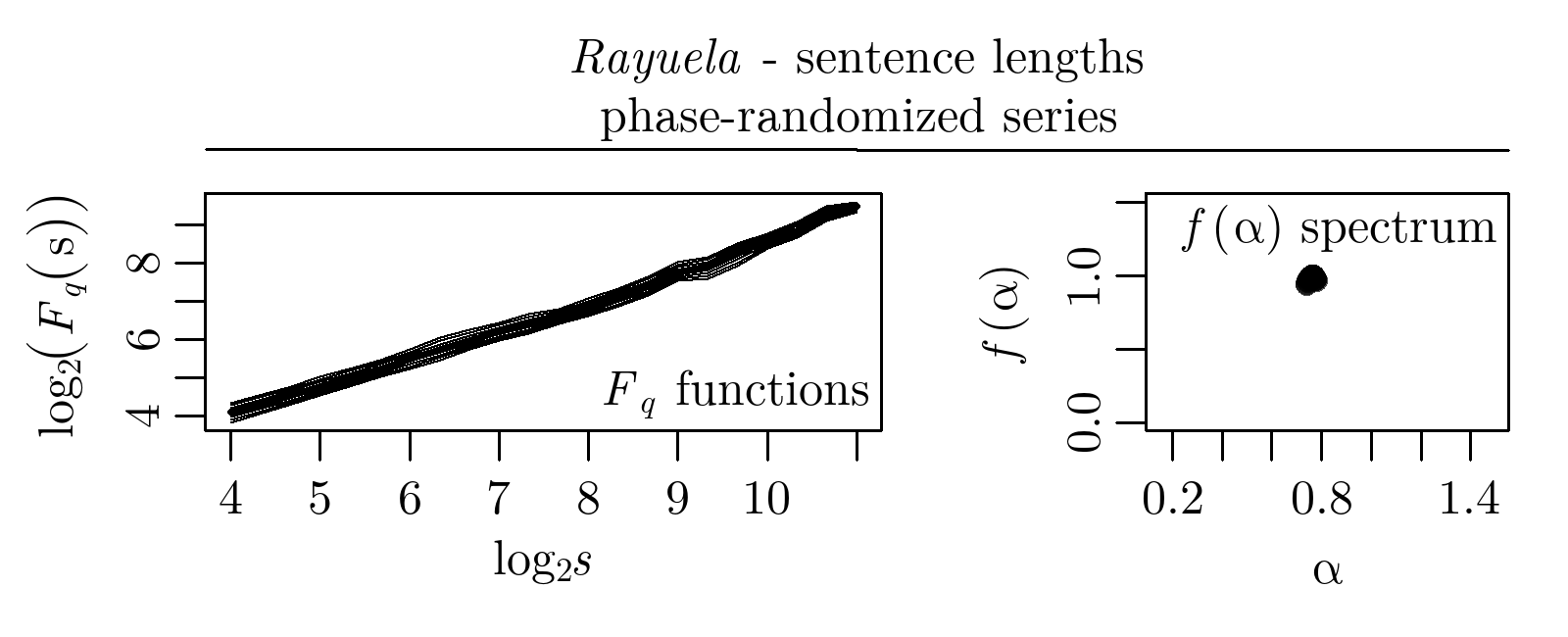}

\includegraphics[width=\SingleSurrogatePlotWidth]{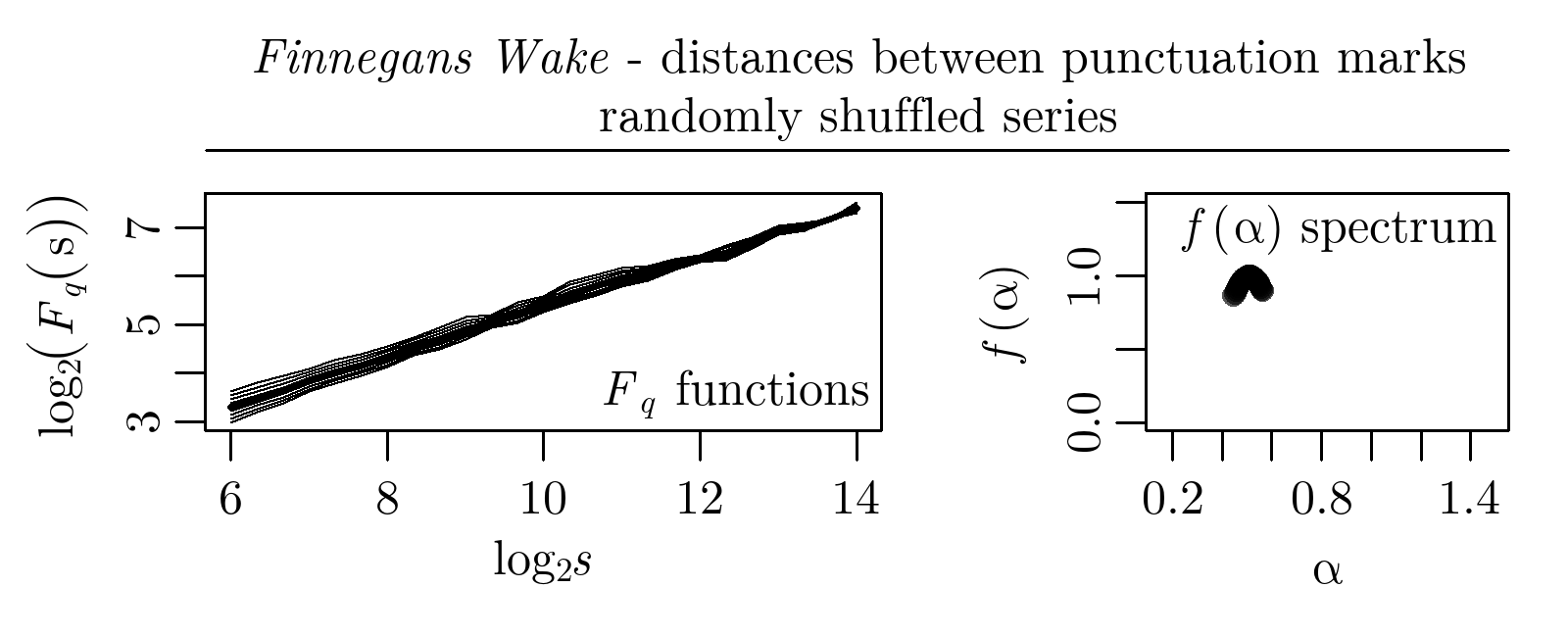}
\hfill
\includegraphics[width=\SingleSurrogatePlotWidth]{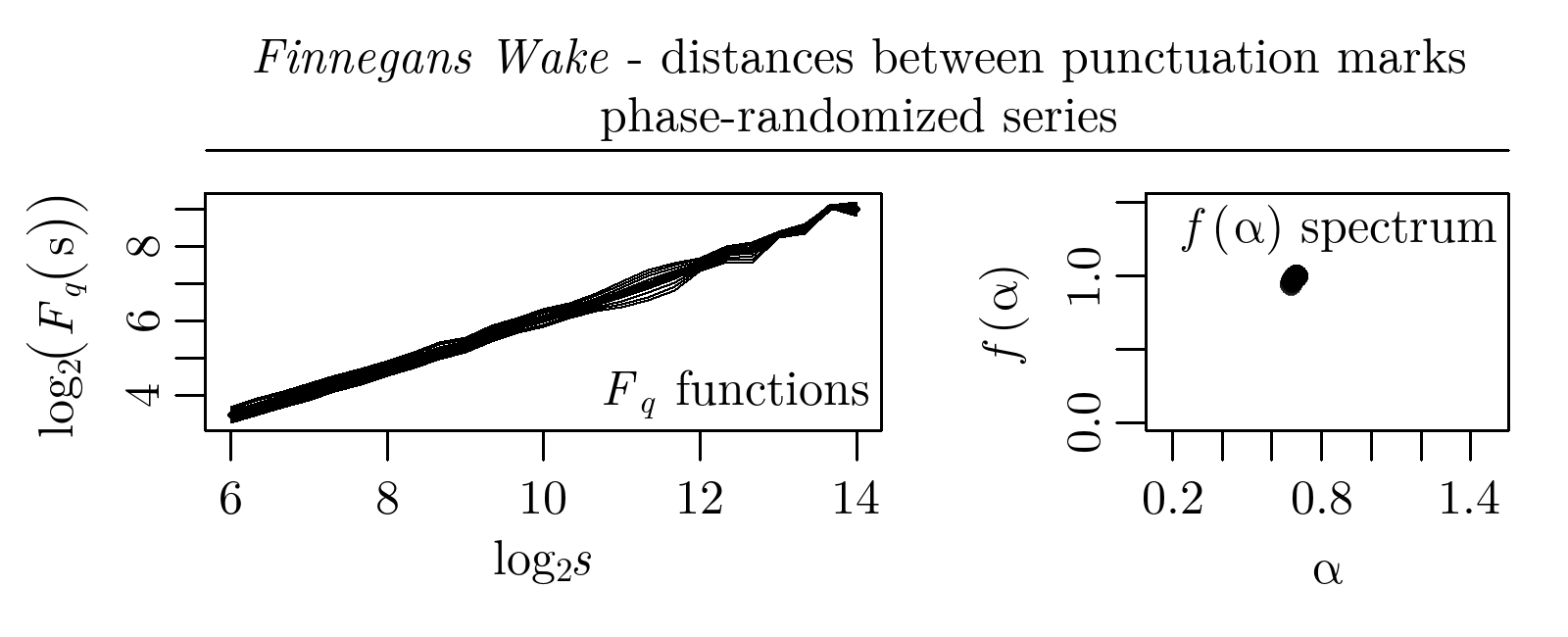}

\includegraphics[width=\SingleSurrogatePlotWidth]{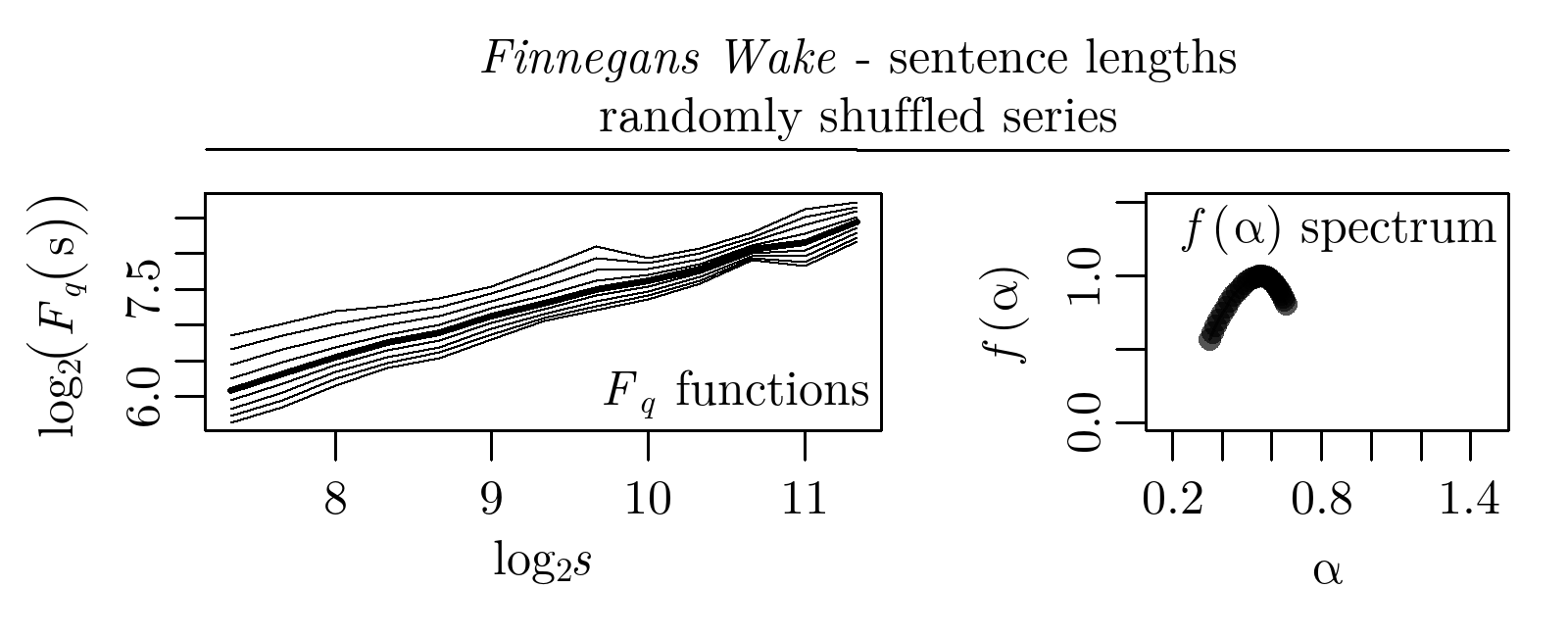}
\hfill
\includegraphics[width=\SingleSurrogatePlotWidth]{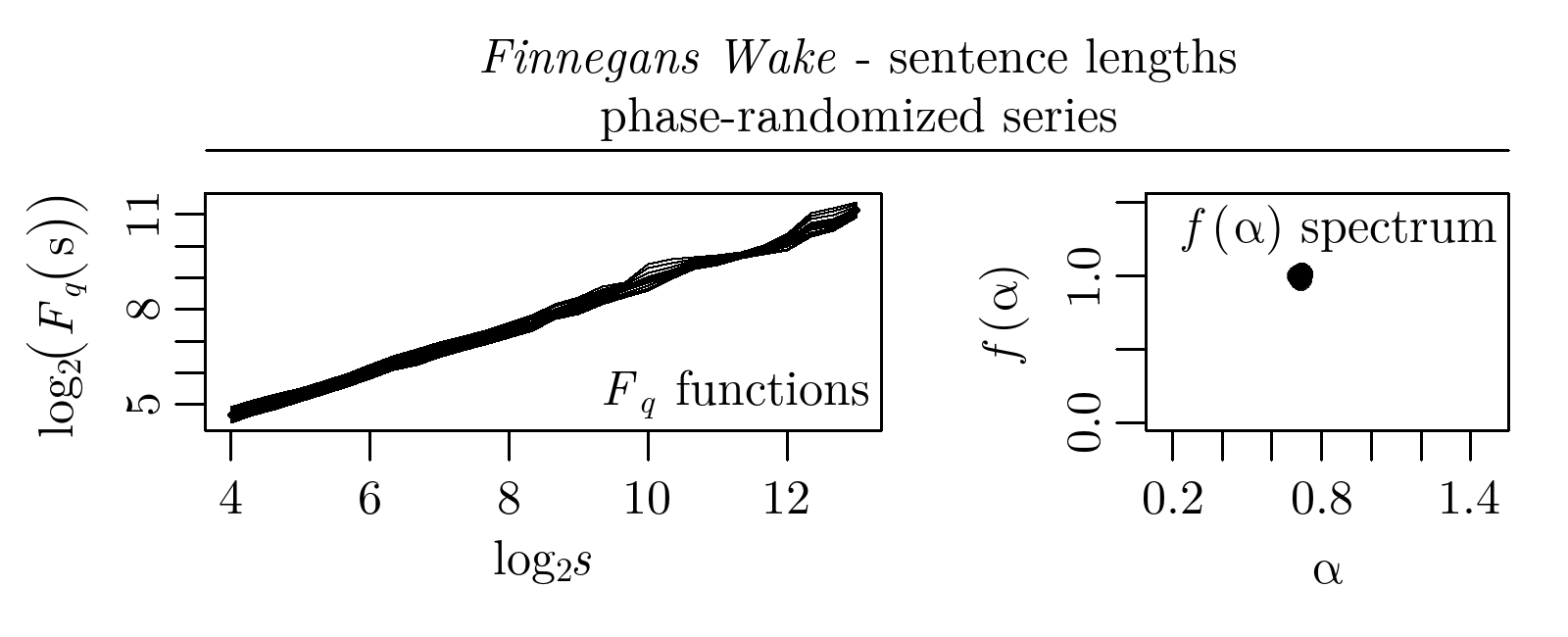}

\includegraphics[width=\SingleSurrogatePlotWidth]{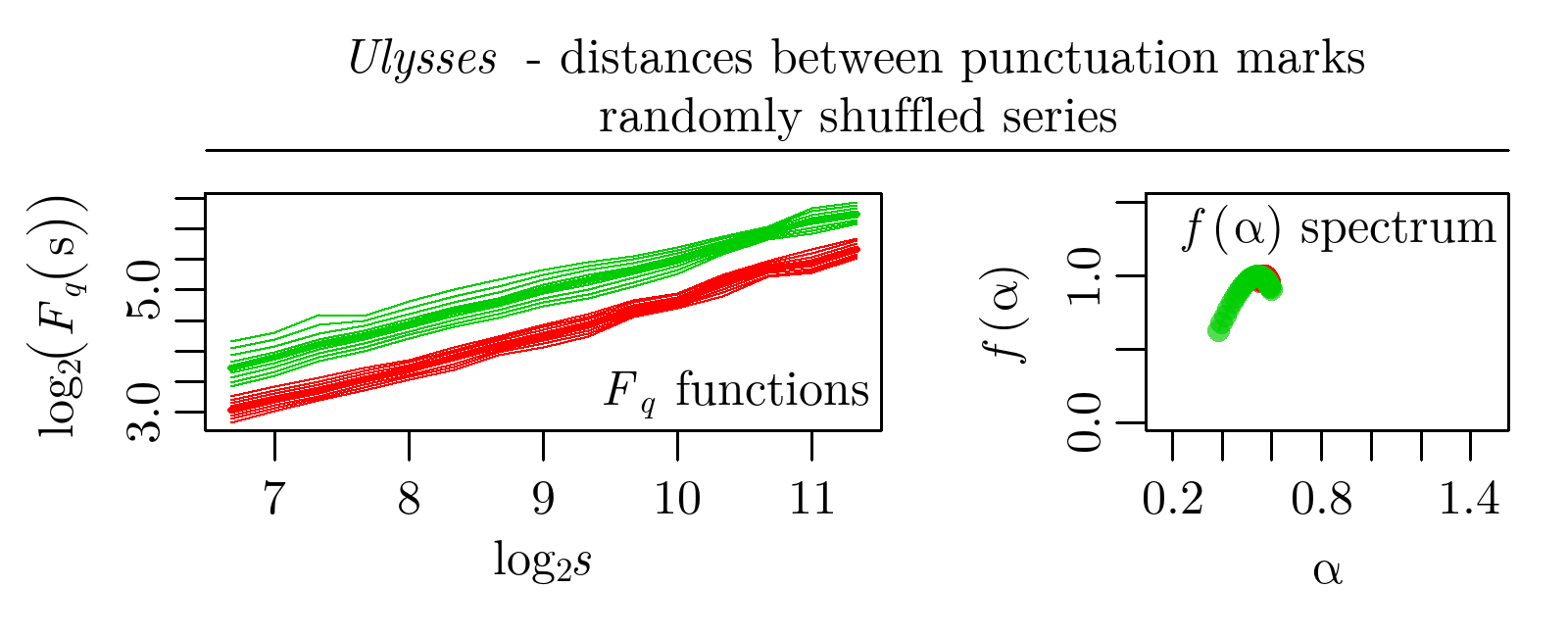}
\hfill
\includegraphics[width=\SingleSurrogatePlotWidth]{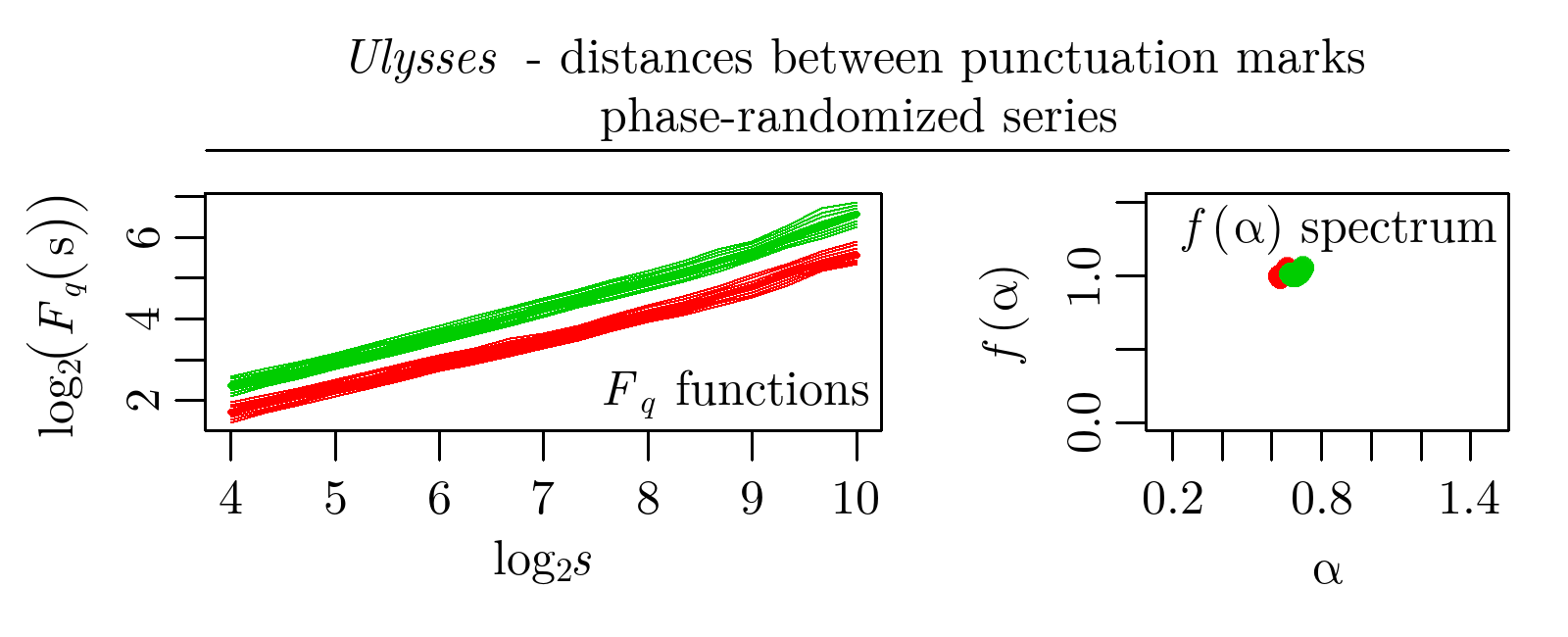}

\includegraphics[width=\SingleSurrogatePlotWidth]{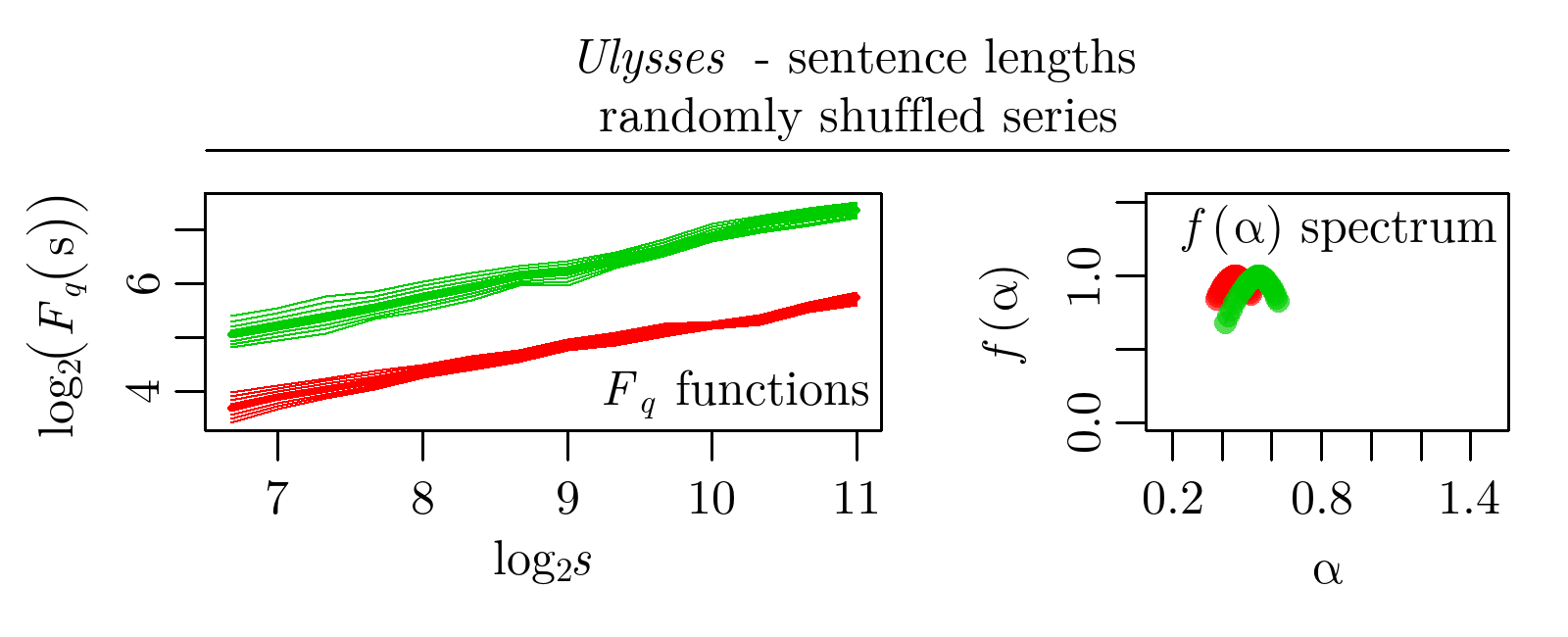}
\hfill
\includegraphics[width=\SingleSurrogatePlotWidth]{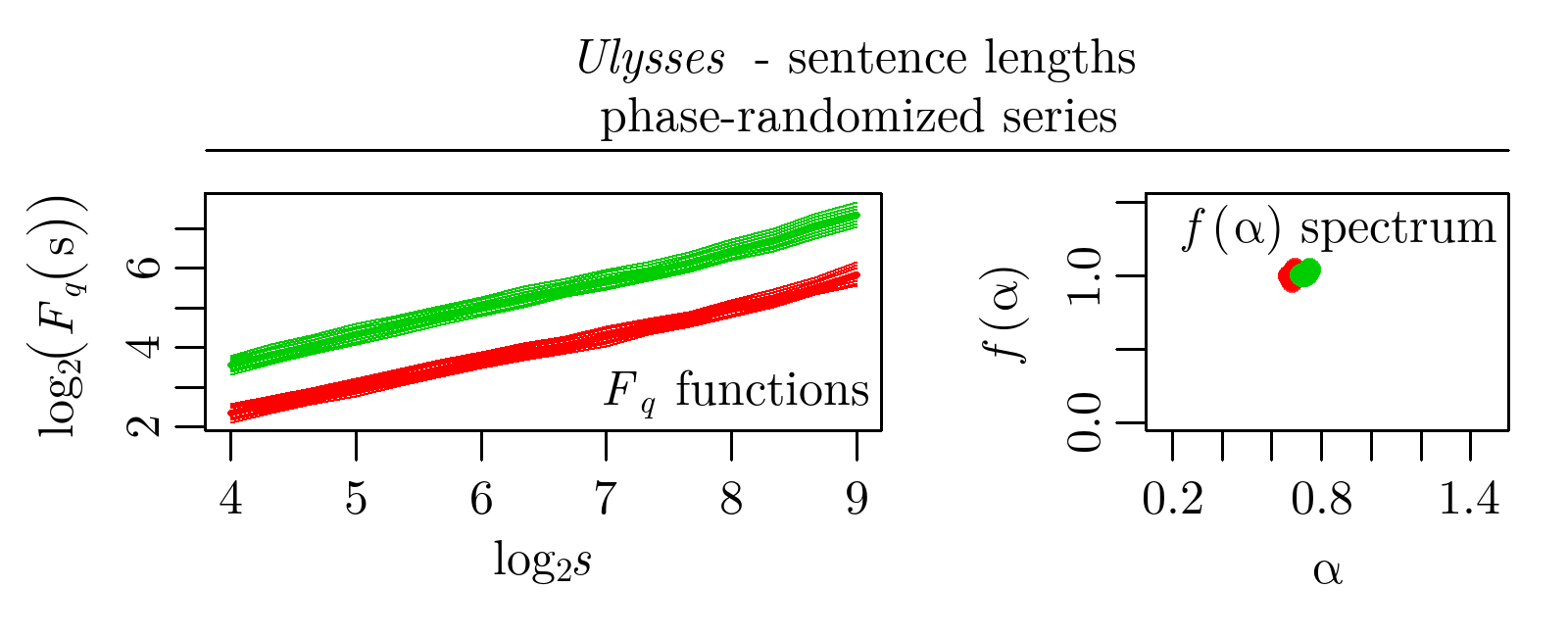}

\caption{Fluctuation functions and singularity spectra obtained by applying MFDFA to exemplary random surrogate time series constructed from the books \textit{Rayuela}, \textit{Finnegans Wake}, and \textit{Ulysses}. For each book, there are two types of base series (sentence lengths, distances between consecutive punctuation marks), and two types of randomization (randomly shuffled series, phase-randomized series). As in Figs.~\ref{fig:MFDFA_allpunctuation} and \ref{fig:MFDFA_sentences}, the two parts of \textit{Ulysses} are considered separately (plotted in red and green, respectively).}
\label{fig:MFDFA_surrogates}
\end{figure*}

Figure~\ref{fig:MFDFA_surrogates} shows the results of MFDFA applied to exemplary random surrogate series, constructed from the series representing distances between consecutive punctuation marks and sentence lengths in the books \textit{Rayuela}, \textit{Finnegans Wake}, and \textit{Ulysses}.

\FloatBarrier

\section*{References}

\vspace*{-1.5em}

\bibliography{bibliography_bibtex_file}
\vspace*{2.5em}

\end{document}